\RequirePackage{fix-cm}

\documentclass{article}

\usepackage{graphicx}
\usepackage{amssymb}
\usepackage{amsmath}
\usepackage{eucal}

\usepackage[driverfallback=dvipdfm,breaklinks=true,pdfcreator={LaTeX e Dvips}]{hyperref}
\usepackage{float}

\usepackage{mathptmx}
\usepackage[numbers]{natbib}
\usepackage{color}
\usepackage{ulem}
\usepackage{booktabs}
\usepackage{multirow}
\usepackage{textcomp}
\usepackage{bm}

\makeatletter
\let\MYcaption\@makecaption
\makeatother
\usepackage{subcaption}
      \makeatletter
\let\@makecaption\MYcaption
\makeatother

\usepackage[ruled,vlined]{algorithm2e}

\usepackage{placeins}

\title{4D Topology optimization of moving rigid bodies in fluid flows}

\author{Yuta Tanabe$^\text{a,}\footnote{Corresponding author: {\tt 4524702@ed.tus.ac.jp} (Yuta Tanabe)}$,
        Kentaro Yaji$^\text{b}$ ,
        Kuniharu Ushijima$^\text{a}$ \\[12pt]
$^\text{a}$\textit{Department of Mechanical Engineering, Tokyo University of Science,}\\
              \textit{6-3-1, Niijuku,
              Katsushika-ku, Tokyo 125-8585, Japan}\\
$^\text{b}$\textit{Department of Mechanical Engineering, The University of Osaka,}\\
              \textit{2-1, Yamadaoka,
              Suita, Osaka 565-0871, Japan}}

\begin{document}

\maketitle

\begin{abstract}
    This study applies \textit{4D topology optimization}, a framework for simultaneously optimizing the morphology and motion of a system, to a rigid body that induces fluid flow.
    The rigid body shape is represented on a design grid that is independent of the analysis grid, and at each time step, it undergoes rigid-body motion before being mapped onto the analysis grid.
    The shape is represented using a pseudo-density method, while the motion is directly parametrized by the positions at discrete time steps and smoothed using a temporal filtering technique.
    The fluid dynamics are evaluated through the lattice kinetic scheme, an extended version of the lattice Boltzmann method.
    Design sensitivities with respect to both shape and motion are derived via the adjoint variable method, and the shape and motion are intentionally updated simultaneously during the optimization process.
    Finally, two- and three-dimensional numerical examples are presented and discussed from a physical perspective. 
    Furthermore, the effectiveness of the proposed method is demonstrated by comparison with cases in which only the shape or motion is optimized, as well as through several parameter studies.
\end{abstract}

\section*{Graphical abstract}
\begin{center}
    \includegraphics[width=0.9\columnwidth]{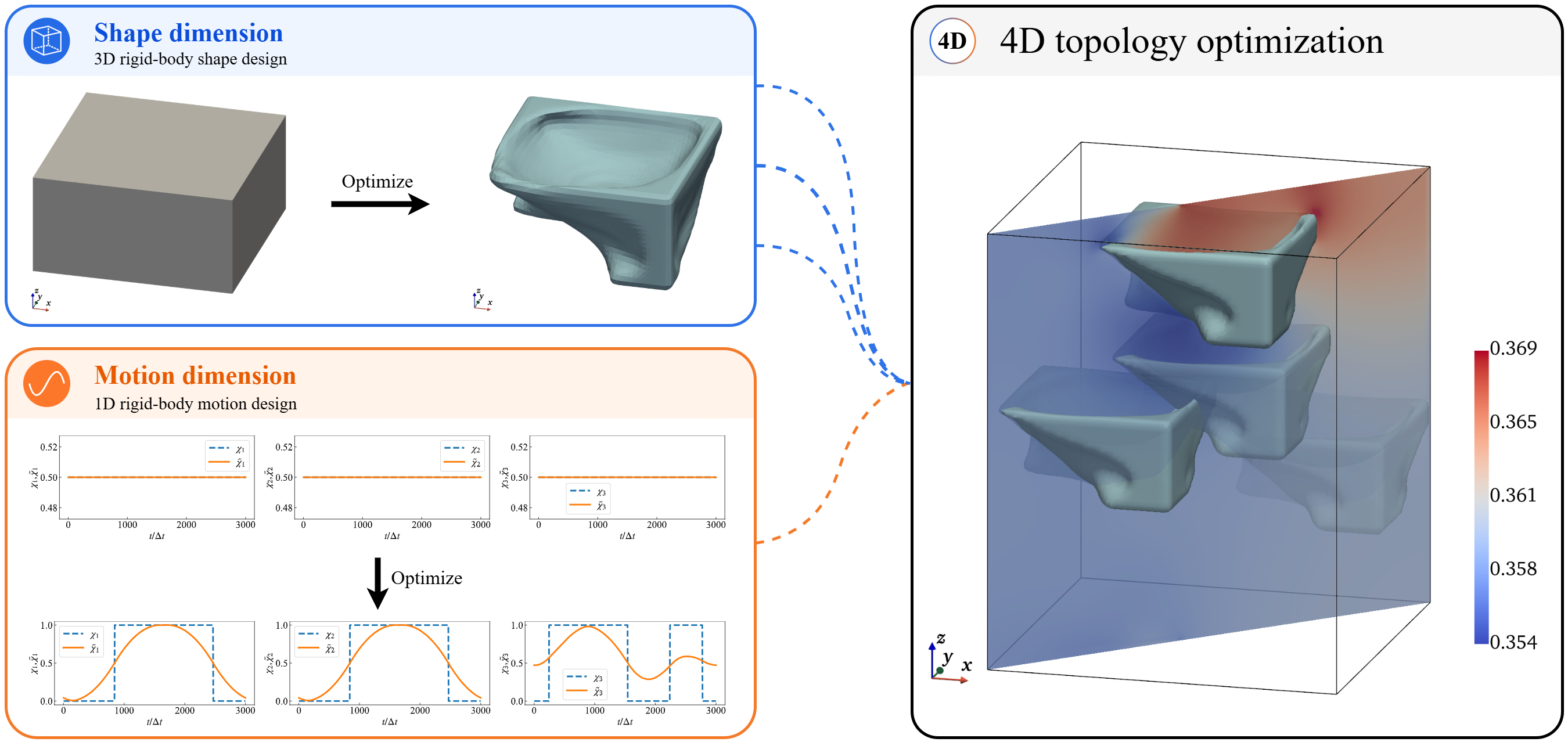}
\end{center}

\section{Introduction}
\label{sec1}

The shape and motion of a mechanical device are inseparably linked in optimizing its performance.
In this study, we focus on a rigid body device operating in fluid flow and develop a method for the simultaneous optimization of its shape and motion.

In the context of structural optimization, topology optimization is one of the most powerful methods due to its high representational capacity.
Traditional optimization approaches, such as size optimization and shape optimization, are limited in their abilities to induce significant geometric changes, whereas topology optimization can achieve such drastic transformations.
Topology optimization was originally introduced into structural mechanics optimization by Bends{\o}e and Kikuchi~\cite{bendsoe1988generating}.
Since then, it has been widely applied to various physical fields, including fluid flow.
In the field of fluid flow, Borrvall and Petersson first applied topology optimization to Stokes flow~\cite{borrvall2003topology}, and subsequent studies extended its application to Navier--Stokes flow~\cite{gersborg2005topology,Olesen2006}, turbulent flow~\cite{kontoleontos2013adjoint,DILGEN2018363}, unsteady flow~\cite{kreissl2011topology,DENG20116688}, and both forced convection~\cite{yoon2010topological,Matsumori2013} and natural convection~\cite{alexandersen2014topology,ALEXANDERSEN2016876}.

The existence studies of topology optimization for the fluid flow problems mentioned above mostly consider stationary components; however, in realistic cases, many kinds of devices move in fluid flow.
Accordingly, topology optimization related to fluid flow problems has also been applied to movable components, such as turbines.
Romero and Silva proposed a topology optimization method for rotating machinery based on steady-state analysis in a rotating coordinate system~\cite{ROMERO2014268}, and their method has been applied to a variety of problem settings, such as 3D problems~\cite{OKUBO202116}, turbulent flow~\cite{SA2021113551,Alonso2022}, and non-Newtonian flow~\cite{Romero2017}.
Their methods have also been applied to the simultaneous design of movable and stationary components, such as turbine blades and guide vanes~\cite{ALONSO2023592,ALONSO2025115982}.
Recently, the method has been applied to flow-rate maximization problems~\cite{SASAKI2026116599}.

Although the method of performing steady-state analysis in a rotating coordinate system is an effective choice in terms of computational cost, it requires the geometrical setting to be isotropic; therefore, it is necessary to introduce a direct representation of the motion of the moving device in unsteady analysis for general motion.
The approach tracking the device motion directly requires the efficient unsteady fluid solver.
Various unsteady fluid topology optimization methods have been reported, such as those based on the Finite Element Method (FEM)~\cite{kreissl2011topology,DENG20116688}, Finite Volume Method (FVM)~\cite{Katsumata2025Topology}, and Lattice Boltzmann Method (LBM)~\cite{NORGAARD2016291}.
In particular, LBM is suitable for unsteady problems because of its fully explicit scheme.
In this study, we use the Lattice Kinetic Scheme (LKS)~\cite{Coveney2002lattice}, which is an extended version of LBM~\cite{kruger2017lattice} that enables more stable computations while reducing memory consumption.

As an approach for applying topology optimization methods to moving objects in fluid while explicitly representing their motion, we proposed the grid separation approach~\cite{TANABE2025114620}.
In this approach, the design grid, in which design variables are defined, is separated from the analysis grid, in which the state fields are computed.
At each time step, the design grid undergoes rigid-body motion and overlaps the analysis grid, and quantities used in the analysis are transferred between the grids.
Also, the rigid-body motion is represented explicitly and directly; therefore, arbitrary motion can be represented.
Recently, this approach has been applied to passive motion, in which the rigid body is driven by fluid flow forces~\cite{tanabe2026topologyoptimizationpassivelymoving}.

In the previously mentioned studies dealing with moving objects in fluid, the motion is either prescribed, such as simple rotation or translation, or not explicitly controllable, such as motion computed from equations of motion.
In contrast, we focus on the simultaneous optimization of shape and motion; in other words, additional design variables representing the trajectory of the rigid body are considered together with the shape-related design variables.
The motion is represented by the position or rotation angle at each time step, and temporal smoothing is applied using a filter.
This optimization setting, which optimizes both structure and motion simultaneously, is called \textit{4D topology optimization}.
The method was proposed in the field of structural optimization~\cite{YUHN2023116187}, and impressive results, such as running or jumping soft robots, were demonstrated.
In that work, actuation signals were optimized together with the shape as additional design variables in the temporal domain.
To the best of the authors' knowledge, topology optimization for fluid flow problems with motion control has not yet been studied, although fluid topology optimization combined with boundary or source-term controls has been investigated~\cite{DENG2014374,Deng2018}.

The shape is represented using the pseudo-density method~\cite{bendsoe2003topology}.
The design sensitivity is computed using the Adjoint Lattice Kinetic Scheme (ALKS)~\cite{TANABE2025114001}, which combines the LKS with the adjoint variable method.
In the original ALKS, the design sensitivity is derived only for the stationary components; therefore, we additionally derived for the design sensitivity for the motion.

This paper is organized as follows.
Section~\ref{sec2} introduces the fundamental equations of the proposed method.
Section~\ref{sec3} explains the details of the numerical implementation.
Section~\ref{sec4} presents numerical examples for two- and three-dimensional problems.
Finally, Section~\ref{sec5} concludes this study. 

\section{Formulation}
\label{sec2}

\subsection{4D topology optimization for moving rigid body in fluid}
\label{sec21}

Topology optimization is one of the structural optimization methods.
The basic concept of this method is to reformulate a structural optimization problem as a material distribution optimization problem; that is, whether solid or fluid material should be placed at each point in a fixed design domain is determined so as to minimize or maximize an objective function subject to given constraints.
Additionally, in \textit{4D topology optimization}, the temporal domain variables are also optimized simultaneously.
The optimization problem is formulated as follows:
\begin{align}
    \begin{array}{lll}
        \underset{\gamma,\bm{\chi},\psi}{\text{maximize }} & J\left(\gamma,\bm{\chi},\psi,U\left(\gamma,\bm{\chi},\psi\right)\right)                                                                \\
        \text{subject to}                                  & G_j\left(\gamma,\bm{\chi},\psi,U\left(\gamma,\bm{\chi},\psi\right)\right)\leq0\vspace{1.5mm} & \left(j=1,2,\cdots,N_\text{cons}\right) \\
                                                           & 0\leq\gamma\left(\bm{\xi}\right)\leq1                                                                                                  \\
                                                           & \bm{0}\leq\bm{\chi}\left(t\right)\leq\bm{1}                                                                                            \\
                                                           & 0\leq\psi\left(t\right)\leq1,
    \end{array} \label{eq:optimization_formulation}
\end{align}
where $J$ and $G_j$ denote the objective functional and the $j$-th constraint functional, respectively, and $N_\text{cons}$ represents the number of constraint functionals.
The vector $\bm{\xi}$ denotes the local coordinates in the design domain.
Here, $\bm{\xi}=\left(\xi_1,\xi_2\right)^\text{T}$ for two-dimensional problems, whereas $\bm{\xi}=\left(\xi_1,\xi_2,\xi_3\right)^\text{T}$ for three-dimensional problems.
The local coordinate axes are denoted by $\xi$, $\eta$, and $\zeta$, and $\xi_1$, $\xi_2$, and $\xi_3$ correspond to these directions, respectively.
It should be noted that, in general, the design domain is fixed with respect to the analysis domain, on which the state fields are computed.
In contrast, in the present study, the design domain undergoes rigid-body motion and changes its position over time.
Therefore, the design variable $\gamma$ varies with time when observed in the analysis domain, while it remains unchanged in the design domain.
The details of the relationship between the design and analysis domains are described in Section~\ref{sec23}.
The design variable $\gamma$ represents whether solid or fluid material is assigned at each point and takes continuous values between $0$ and $1$.
This shape representation method is called the pseudo-density method~\cite{bendsoe2003topology}.
Here, $\gamma=0$ corresponds to fluid, $\gamma=1$ corresponds to solid, and intermediate values represent porous media.
$\bm{\chi}$ and $\psi$ are design variables defined in the temporal domain and correspond to translational and rotational motions, respectively, whose details are described in Section~\ref{sec24}.
Here, $\bm{\chi}=\left(\chi_1,\chi_2\right)^\text{T}$ for two-dimensional problems, whereas $\bm{\chi}=\left(\chi_1,\chi_2,\chi_3\right)^\text{T}$ for three-dimensional problems, where the first, second, and third components correspond to the $x$-, $y$-, and $z$-directions, respectively.
The bounds on $\bm{\chi}$ in Eq.~\eqref{eq:optimization_formulation} are imposed componentwise.
The state variable $U$ depends implicitly on $\gamma$, $\bm{\chi}$ and $\psi$, and both the objective functional $J$ and the constraint functionals $G_j$ depend on $\gamma$, $\bm{\chi}$, $\psi$ and $U$.

For fluid flow problems, the state variables $U$ consist of the pressure $p$ and the velocity components $u_\alpha$, which are governed by the continuity equation and the momentum conservation equation given by
\begin{align}
     & \frac{\partial u_\alpha}{\partial x_\alpha}=0,                                                                                                                                                                               \\
     & \frac{\partial u_\alpha}{\partial t}+u_\beta\frac{\partial u_\alpha}{\partial x_\beta}=-\frac{\partial p}{\partial x_\alpha}+\nu\frac{\partial^2u_\alpha}{\partial x_\beta^2}-\kappa\left(u_\alpha-u_\alpha^\text{S}\right).
\end{align}
Here, the subscripts $\alpha$ and $\beta$ denote the spatial directions, with $\alpha,\beta=1,2$ for two-dimensional problems and $\alpha,\beta=1,2,3$ for three-dimensional problems.
The indices $1$, $2$, and $3$ correspond to the $x$-, $y$-, and $z$-directions, respectively.
The parameters $\kappa$ and $u_\alpha^\text{S}$ represent the Brinkman coefficient and the rigid-body velocity at each point in the design domain, respectively.
The Brinkman coefficient $\kappa$ takes a large value only in the solid region, such that the fluid velocity approaches the rigid-body velocity in that region.
Although $\kappa$ depends on the design variable $\gamma$, its specific formulation is described in Section~\ref{sec23}.
The parameter $\nu$ denotes the kinematic viscosity.

\subsection{Lattice Kinetic Scheme}
\label{sec22}

\begin{figure}[t]
    \centering
    \begin{minipage}[t]{0.49\columnwidth}
        \centering
        \includegraphics[width=\columnwidth]{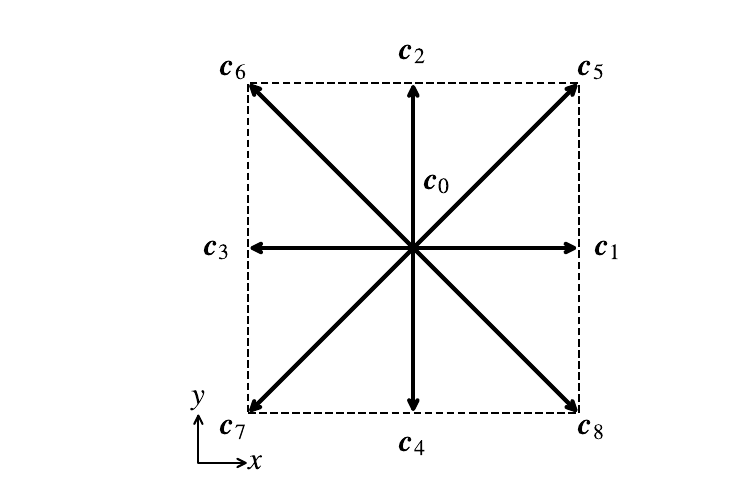}
        \subcaption{D2Q9}
        \label{fig:d2q9}
    \end{minipage}
    \begin{minipage}[t]{0.49\columnwidth}
        \centering
        \includegraphics[width=\columnwidth]{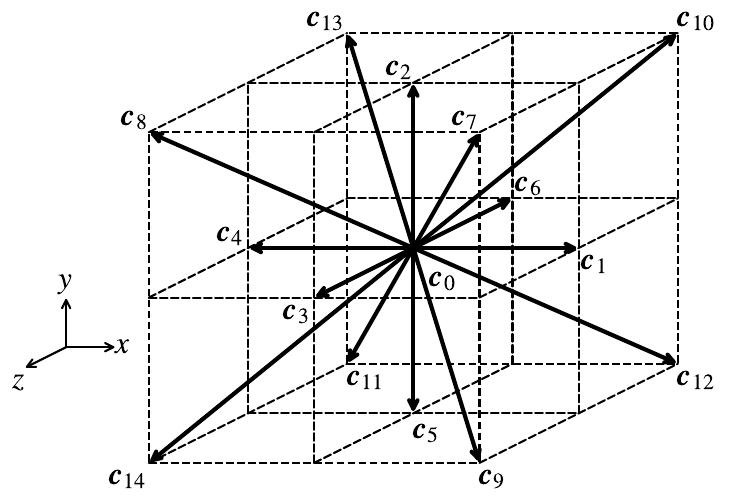}
        \subcaption{D3Q15}
        \label{fig:d3q15}
    \end{minipage}
    \caption{Velocity sets for each lattice gas model.
        Reproduced from Ref.~\citep{TANABE2025114620} under the CC-BY license.}
\end{figure}

We compute the fluid state variables using the Lattice Kinetic Scheme (LKS)~\cite{Coveney2002lattice}, which is an extended version of the Lattice Boltzmann Method (LBM)~\cite{kruger2017lattice}.
In LBM-family methods, the fluid is modeled by fictitious particles with a finite set of discrete velocities.
The particle distribution functions $f_i$ are governed by the discrete-velocity Boltzmann equation given by
\begin{equation}
    \text{Sh}\frac{\partial f_i}{\partial t}+c_{i\alpha}\frac{\partial f_i}{\partial x_\alpha}=-\frac{1}{\varepsilon}\left(f_i-f_i^\text{eq}\right)-3w_ic_{i\alpha}\kappa\left(u_\alpha-u_\alpha^\text{S}\right), \label{eq:discrete_velocity_boltzmann}
\end{equation}
where the subscript $i$ denotes the particle associated with the $i$-th discrete velocity direction.
The parameters $w_i$ and $c_{i\alpha}$ represent the weight and the particle velocity vector, respectively, and depend on the lattice gas model.
In this study, the D2Q9 model (see Fig.~\ref{fig:d2q9}) is employed for two-dimensional cases, whereas the D3Q15 model (see Fig.~\ref{fig:d3q15}) is used for three-dimensional cases~\cite{kruger2017lattice}.
The parameters $\text{Sh}$ and $\varepsilon$ denote the Strouhal number and a dimensionless parameter of the same order as the Knudsen number, respectively.
The function $f_i^\text{eq}$ is the local equilibrium distribution function, whose specific form is described in the next paragraph.
Eq.~\eqref{eq:discrete_velocity_boltzmann} is solved in time using the following discretized form:
\begin{align}
     & f_i^*\left(\bm{x}+\bm{c}_i\Delta x,t+\Delta t\right)=f_i\left(\bm{x},t\right)-\frac{1}{\tau}\left\{f_i\left(\bm{x},t\right)-f_i^\text{eq}\left(\bm{x},t\right)\right\}, \label{eq:lattice_boltzmann_equation}                       \\
     & f_i\left(\bm{x},t\right)=f_i^*\left(\bm{x},t\right)-3\Delta xw_ic_{i\alpha}\kappa\left(\bm{x},t\right)\left\{u_\alpha\left(\bm{x},t\right)-u_\alpha^\text{S}\left(\bm{x},t\right)\right\}, \label{eq:lattice_boltzmann_equation_ex}
\end{align}
where $\Delta x$ and $\Delta t$ denote the grid spacing and the time step, respectively.
$\tau$ is the dimensionless relaxation time parameter, and is given by $\tau=\varepsilon/\Delta x$.

Since the parameter $\tau$ is set to unity in the LKS, Eq.~\eqref{eq:lattice_boltzmann_equation} can be simplified as
\begin{equation}
    f_i^*\left(\bm{x},t\right)=f_i^\text{eq}\left(\bm{x}-\bm{c}_{i\alpha}\Delta x,t-\Delta t\right).
\end{equation}
In this study, the equilibrium distribution function $f_i^\text{eq}$ is adopted based on the formulation proposed by Inamuro~\citep{Coveney2002lattice}, and is expressed as
\begin{equation}
    f_i^\text{eq}=w_i\left\{3p+3c_{i\alpha}u_\alpha+\frac{9}{2}\left(c_{i\alpha}u_\alpha\right)^2-\frac{3}{2}u_\alpha^2\right\}+\Delta xw_iA\left(\frac{\partial u_\beta}{\partial x_\alpha}+\frac{\partial u_\alpha}{\partial x_\beta}\right)c_{i\alpha}c_{i\beta},
\end{equation}
where, $A$ is a parameter related to the kinematic viscosity $\nu$, and is given by
\begin{equation}
    A=\frac{3}{4}-\frac{9}{2}\frac{\nu}{\Delta x}.
\end{equation}
The macroscopic variables, such as the pressure $p$ and the velocity components $u_\alpha$, are obtained as moment of $f_i$:
\begin{align}
     & p^*\left(\bm{x},t\right)=\frac{1}{3}\sum_{i=0}^{Q-1}f_i^\text{eq}\left(\bm{x}-\bm{c}_i\Delta x,t-\Delta t\right), \label{eq:p_tmp}         \\
     & u_\alpha^*\left(\bm{x},t\right)=\sum_{i=0}^{Q-1}c_{i\alpha}f_i^\text{eq}\left(\bm{x}-\bm{c}_i\Delta x, t-\Delta t\right). \label{eq:u_tmp}
\end{align}
Here, quantities with the superscript ``*'' represent intermediate values without the effect of the external force term.
The corresponding macroscopic variables including the force effects are computed from the moments of Eq.~\eqref{eq:lattice_boltzmann_equation_ex} as
\begin{align}
     & p\left(\bm{x},t\right)=p^*\left(\bm{x},t\right), \label{eq:p}                                                                                                                                       \\
     & u_\alpha\left(\bm{x},t\right)=\frac{u_\alpha^*\left(\bm{x},t\right)+\Delta x\kappa\left(\bm{x},t\right)u_\alpha^\text{S}\left(\bm{x},t\right)}{1+\Delta x\kappa\left(\bm{x},t\right)}. \label{eq:u}
\end{align}
As shown in Eqs.~\eqref{eq:p_tmp}--\eqref{eq:u}, the LKS requires only the macroscopic variables to be updated in time.
Therefore, for the boundary conditions, $p$ and $u_\alpha$ are directly imposed.

\subsection{Grid separation approach}
\label{sec23}

We employ the grid separation approach~\cite{TANABE2025114620} to represent rigid body moving in a fluid flow.
The basic concept of this method is to separate the design grid, on which the design variables are defined, from the analysis grid, on which the state fields are computed.
At each time step, the design grid undergoes rigid-body motion and is overlapped onto the analysis grid.
It should be noted that, in the actual implementation, not only the design region but also the surrounding margin region undergoes rigid-body motion.
The margin region has a width of several grid points and is entirely filled with fluid.
Quantities required for the time-marching computation, such as the Brinkman coefficient $\kappa$ and the rigid-body velocity at each point $u_\alpha^\text{S}$, are transferred from a grid to the other grid as follows:
\begin{align}
     & \kappa\left(\bm{x},t\right)=\int_DW\left(\bm{x},\bm{x}^\text{ref}\left(\bm{\xi},t\right)\right)\kappa^\text{ref}\left(\bm{\xi}\right)d\Omega^\text{ref}, \label{eq:kappa}                \\
     & u_\alpha^\text{S}\left(\bm{x},t\right)=\int_DW\left(\bm{x},\bm{x}^\text{ref}\left(\bm{\xi},t\right)\right)u_\alpha^\text{S,ref}\left(\bm{\xi},t\right)d\Omega^\text{ref}, \label{eq:u_s}
\end{align}
where quantities with the superscript ``ref'' are defined on the design grid; that is, they are expressed as functions of the local coordinates $\bm{\xi}$.
The specific form of $\bm{x}^\text{ref}$, $\kappa^\text{ref}$, and $u_\alpha^\text{S,ref}$ are described later.
The weight function $W$ is defined as
\begin{align}
     & W\left(\bm{x},\bm{x}^\text{ref}\right)=\prod_{\alpha=1}^{d}w\left(x_\alpha-x_\alpha^\text{ref}\right),                   \\
     & w\left(r\right)=\left\{\begin{array}{ll}
                                  \frac{1}{4\Delta x}\left\{1+\cos\left(\frac{\pi r}{2}\right)\right\} & \left|r\right|\le2\Delta x \\
                                  0                                                                    & \left|r\right|>2\Delta x
                              \end{array}\right.,
\end{align}
where $d$ denotes the spatial dimension, with $d=2$ for two-dimensional problems and $d=3$ for three-dimensional problems.
The coordinates $x_1$, $x_2$, and $x_3$ correspond to $x$, $y$ and $z$, respectively.

$\kappa^\text{ref}$ depends on the design variable $\gamma$ and is given by
\begin{equation}
    \kappa^\text{ref}\left(\gamma\right)=\kappa_\text{max}^\text{ref}\frac{q\gamma}{\left(1-\gamma\right)+q}, \label{eq:kappa_ref}
\end{equation}
where $\kappa_\text{max}^\text{ref}$ represents the inverse permeability in the solid region, and $q$ is a convexity parameter.
Their specific values are provided in the numerical examples in Section~\ref{sec4}.
$\bm{x}^\text{ref}$ is obtained through the rigid-body motion as
\begin{equation}
    \bm{x}^\text{ref}=\left(\begin{array}{cc}\cos\theta\left(t\right) & -\sin\theta\left(t\right) \\
             \sin\theta\left(t\right)     & \cos\theta\left(t\right)\end{array}\right)\left(\begin{array}{c}\xi_1-\xi_1^\text{G}\\\xi_2-\xi_2^\text{G}\end{array}\right)+\left(\begin{array}{c}x_1^\text{G}\left(t\right)\\x_2^\text{G}\left(t\right)\end{array}\right), \label{eq:rigid_body_position}
\end{equation}
where $\theta$ and $x_\alpha^\text{G}$ denote the rotation angle of the rigid body and the coordinates of its center of gravity on the analysis grid, respectively.
$\bm{\xi}^\text{G}$ represents the center-of-gravity coordinates on the design grid.
The rigid-body velocity at each point, $u_\alpha^\text{S,ref}$, is obtained by differentiating Eq.~\eqref{eq:rigid_body_position} with respect to $t$:
\begin{equation}
    \bm{u}^\text{S,ref}=\left(\begin{array}{cc}-\sin\theta\left(t\right) & -\cos\theta\left(t\right) \\
             \cos\theta\left(t\right)      & -\sin\theta\left(t\right)\end{array}\right)\left(\begin{array}{c}\xi_1-\xi_1^\text{G}\\\xi_2-\xi_2^\text{G}\end{array}\right)\omega\left(t\right)+\left(\begin{array}{c}u_1^\text{G}\left(t\right)\\u_2^\text{G}\left(t\right)\end{array}\right), \label{eq:rigid_body_velocity}
\end{equation}
where $\omega=d\theta/dt$ is the angular velocity and $u_\alpha^\text{G}=dx_\alpha^\text{G}/dt$ denotes the translational velocity of the center of gravity.

\subsection{Motion representation}
\label{sec24}

We parametrize the rigid-body motion and optimize its parameters simultaneously with the topology optimization.
The coordinates of the rigid-body center of gravity, $x^\text{G}_\alpha$, are parametrized by the design variable $\chi_\alpha\left(t\right)$, which takes continuous values between $0$ and $1$.
First, $\chi_\alpha$ is filtered as follows:
\begin{equation}
    \tilde{\chi}_\alpha\left(t\right)=\frac{\int_\mathcal{I}w_\text{t}\left(t,\tau\right)\chi_\alpha\left(\tau\right)d\tau}{\int_\mathcal{I}w_\text{t}\left(t,\tau\right)d\tau}, \label{eq:chi_filtered}
\end{equation}
where $w_\text{t}$ is the weight function given by
\begin{equation}
    w_\text{t}\left(t,\tau\right)=\left\{\begin{array}{cl}\frac{R_\text{t}-\left|t-\tau\right|}{R_\text{t}}&\left|t-\tau\right|\leq R_\text{t} \\0&\left|t-\tau\right|>R_\text{t}\end{array}\right..
\end{equation}
Here, $R_\text{t}$ denotes the filter radius, and its specific value is provided in the numerical examples in Section~\ref{sec4}.
This filtering scheme is applied because $\chi_\alpha$ is allowed to have a discontinuous distribution; that is, the rigid body may exhibit abrupt positional changes at the next time step if the unfiltered $\chi_\alpha$ is directly employed.
Next, $x^\text{G}_\alpha$ is computed from $\tilde{\chi}_\alpha$ as
\begin{equation}
    x^\text{G}_\alpha\left(t\right)=\left(x_\alpha^\text{max}-x_\alpha^\text{min}\right)\tilde{\chi}_\alpha\left(t\right)+x_\alpha^\text{min}, \label{eq:x_g}
\end{equation}
where $x_\alpha^\text{max}$ and $x_\alpha^\text{min}$ denote the upper and lower bounds of the admissible coordinate range, respectively.
The velocity of the rigid-body center of gravity, $u^\text{G}_\alpha$, is obtained as follows:
\begin{equation}
    u^\text{G}_\alpha=\frac{dx^\text{G}_\alpha}{dt}. \label{eq:u_g}
\end{equation}

In the same manner as the translational motion, the rotation angle $\theta$ is also parametrized by the design variable $\psi$ as follows:
\begin{align}
     & \tilde{\psi}\left(t\right)=\frac{\int_\mathcal{I}w_\omega\left(t,\tau\right)\psi\left(\tau\right)d\tau}{\int_\mathcal{I}w_\omega\left(t,\tau\right)d\tau}, \label{eq:psi_filtered} \\
     & \theta\left(t\right)=\left(\theta^\text{max}-\theta^\text{min}\right)\tilde{\psi}\left(t\right)+\theta^\text{min}. \label{eq:theta}
\end{align}
Here, $w_\omega$ is the weight function and has the same shape as $w_\text{t}$, except that $R_\text{t}$ is replaced by $R_\omega$, the filter radius for $\psi$.
The angular velocity, $\omega$ is given by
\begin{equation}
    \omega=\frac{d\theta}{dt}. \label{eq:omega}
\end{equation}
It should be noted that, in this study, $\theta$ is defined as the rotation angle about an axis parallel to the $z$-axis.
However, this formulation can be readily extended to general three-dimensional rotations by using Euler angles or quaternions.

\subsection{Sensitivity analysis}
\label{sec25}

We describe the sensitivity analysis method for a general functional $J$ based on the continuous adjoint variable method.
$J$ is defined as
\begin{equation}
    J=\int_\mathcal{I}\int_\mathcal{O}J_\mathcal{O}d\Omega dt+\int_\mathcal{I}\int_{\partial\mathcal{O}}J_{\partial\mathcal{O}}d\Gamma dt,
\end{equation}
where $J_\mathcal{O}$ and $J_{\partial\mathcal{O}}$ denote the integrands over the domain and its boundary, respectively.
Using the temporal integrand, $J_\mathcal{I}$, the functional $J$ can also be expressed as
\begin{equation}
    J=\int_\mathcal{I}J_\mathcal{I}dt.
\end{equation}
Following the Adjoint Lattice Kinetic Scheme (ALKS)~\cite{TANABE2025114001}, which combines the continuous adjoint variable method with the Lattice Kinetic Scheme (LKS)~\cite{Coveney2002lattice}, the Lagrangian $L$ is defined as
\begin{align}
    L= & J+\int_\mathcal{I}\int_\mathcal{O}\sum_{i=0}^{Q-1}\tilde{f}_i\left\{\text{Sh}\frac{\partial f_i}{\partial t}+c_{i\alpha}\frac{\partial f_i}{\partial x_\alpha}+\frac{1}{\varepsilon}\left(f_i-f_i^\text{eq}\right)+3w_ic_{i\alpha}\kappa\left(u_\alpha-u_\alpha^\text{S}\right)\right\}d\Omega dt \notag \\
       & +\int_\mathcal{I}\tilde{u}^\text{G}_\alpha\left(u^\text{G}_\alpha-\frac{dx^\text{G}_\alpha}{dt}\right)dt+\int_\mathcal{I}\tilde{\omega}\left(\omega-\frac{d\theta}{dt}\right)dt, \label{eq:lagrangian}
\end{align}
where $\tilde{f}_i$, $\tilde{u}_\alpha^\text{G}$, and $\tilde{\omega}$ are the lagrange multipliers.
The derivative of $L$ with respect to $\gamma$, $\chi_\alpha$, and $\psi$ is rearranged so that the derivatives of the state fields do not explicitly appear.
Consequently, since $\langle\delta J/\delta\gamma,\delta\gamma\rangle=\langle\delta L/\delta\gamma,\delta\gamma\rangle$, $\langle\delta J/\delta\chi_\alpha,\delta\chi_\alpha\rangle=\langle\delta L/\delta\chi_\alpha,\delta\chi_\alpha\rangle$, and $\langle\delta J/\delta\psi,\delta\psi\rangle=\langle\delta L/\delta\psi,\delta\psi\rangle$, the design sensitivity of $J$ is expressed as follows:
\begin{align}
    \left\langle\frac{\delta J}{\delta\gamma},\delta\gamma\right\rangle=           & \int_\mathcal{I}\int_D3\frac{\partial\kappa^\text{ref}}{\partial\gamma}\left(u_\alpha^\text{ref}-u_\alpha^\text{S,ref}\right)\tilde{u}_\alpha^\text{ref}\delta\gamma d\Omega^\text{ref}dt, \label{eq:dJdgamma}                                                  \\
    \left\langle\frac{\delta J}{\delta\chi_\alpha},\delta\chi_\alpha\right\rangle= & \int_\mathcal{I}\int_\mathcal{O}\int_D3\kappa^\text{ref}\frac{\partial W}{\partial x_\alpha^\text{ref}}\delta x_\alpha^\text{G}\left(u_\alpha-u_\alpha^\text{S}\right)\tilde{u}_\alpha d\Omega^\text{ref}d\Omega dt \notag                                      \\
                                                                                   & -\int_\mathcal{I}\int_\mathcal{O}\int_D3\kappa\tilde{u}_\alpha u_\alpha^\text{S,ref}\frac{\partial W}{\partial x_\alpha^\text{ref}}\delta x_\alpha^\text{G}d\Omega^\text{ref}d\Omega dt \notag                                                                  \\
                                                                                   & +\left[\tilde{u}_\alpha^\text{G}\delta x_\alpha^\text{G}\right]_{t_0}^{t_1}-\int_\mathcal{I}\frac{d\tilde{u}_\alpha^\text{G}}{dt}\delta x_\alpha^\text{G}dt, \label{eq:dJdchi}                                                                                  \\
    \left\langle\frac{\delta J}{\delta\psi},\delta\psi\right\rangle=               & \int_\mathcal{I}\int_\mathcal{O}\int_D3\kappa^\text{ref}\frac{\partial W}{\partial x_\beta^\text{ref}}\frac{\partial x_\beta^\text{ref}}{\partial\theta}\left(u_\alpha-u_\alpha^\text{S}\right)\tilde{u}_\alpha\delta\theta d\Omega^\text{ref}d\Omega dt \notag \\
                                                                                   & -\int_\mathcal{I}\int_\mathcal{O}\int_D3\kappa\tilde{u}_\alpha\frac{\partial u_\alpha^\text{S,ref}}{\partial\theta}W\delta\theta d\Omega^\text{ref}d\Omega dt \notag                                                                                            \\
                                                                                   & -\int_\mathcal{I}\int_\mathcal{O}\int_D3\kappa\tilde{u}_\alpha u_\alpha^\text{S,ref}\frac{\partial W}{\partial x_\beta^\text{ref}}\frac{\partial x_\beta^\text{ref}}{\partial\theta}\delta\theta d\Omega^\text{ref}d\Omega dt \notag                            \\
                                                                                   & +\left[\tilde{\omega}\delta\theta\right]_{t_0}^{t_1}-\int_\mathcal{I}\frac{d\tilde{\omega}}{dt}\delta\theta dt. \label{eq:dJdpsi}
\end{align}
Here, from Eqs.~\eqref{eq:rigid_body_position} and \eqref{eq:rigid_body_velocity}, $\partial x_\alpha^\text{ref}/\partial\theta$ and $\partial u_\alpha^\text{ref}/\partial\theta$ are given by
\begin{align}
     & \left(\begin{array}{c}\displaystyle\frac{\partial x_1^\text{ref}}{\partial\theta}\vspace{2pt} \\ \displaystyle\frac{\partial x_2^\text{ref}}{\partial\theta}\end{array}\right)=\left(\begin{array}{cc}-\sin\theta\left(t\right)&-\cos\theta\left(t\right)\\\cos\theta\left(t\right)&-\sin\theta\left(t\right)\end{array}\right)\left(\begin{array}{c}\xi_1-\xi_1^\text{G}\\\xi_2-\xi_2^\text{G}\end{array}\right),                    \\
     & \left(\begin{array}{c}\displaystyle\frac{\partial u_1^\text{ref}}{\partial\theta}\vspace{2pt}\\ \displaystyle\frac{\partial u_2^\text{ref}}{\partial\theta}\end{array}\right)=\left(\begin{array}{cc}-\cos\theta\left(t\right)&\sin\theta\left(t\right)\\-\sin\theta\left(t\right)&-\cos\theta\left(t\right)\end{array}\right)\left(\begin{array}{c}\xi_1-\xi_1^\text{G}\\\xi_2-\xi_2^\text{G}\end{array}\right)\omega\left(t\right).
\end{align}

The adjoint variable $\tilde{u}_\alpha$ is computed using the ALKS~\cite{TANABE2025114001}.
In the ALKS, the macroscopic adjoint variables without source terms are first obtained as
\begin{align}
     & \tilde{p}^*\left(\bm{x},t\right)=\sum_{i=0}^{Q-1}w_i\tilde{f}_i^\text{eq}\left(\bm{x}+\bm{c}_i\Delta x,t+\Delta t\right), \label{eq:ap_tmp}                                  \\
     & \tilde{u}_\alpha^*\left(\bm{x},t\right)=\sum_{i=0}^{Q-1}w_ic_{i\alpha}\tilde{f}_i^\text{eq}\left(\bm{x}+\bm{c}\Delta x,t+\Delta t\right),                                    \\
     & \tilde{s}_{\alpha\beta}^*\left(\bm{x},t\right)=\sum_{i=0}^{Q-1}w_ic_{i\alpha}c_{i\beta}\tilde{f}_i^\text{eq}\left(\bm{x}+\bm{c}\Delta x,t+\Delta t\right), \label{eq:as_tmp}
\end{align}
where $\tilde{f}_i^\text{eq}$ is defined as
\begin{equation}
    \tilde{f}_i^\text{eq}=\tilde{p}+3c_{i\alpha}\left(\tilde{u}_\alpha+3\tilde{s}_{\alpha\beta}u_\beta-\tilde{p}u_\alpha\right)-\Delta xA\frac{\partial}{\partial x_\beta}\left(\tilde{s}_{\alpha\beta}+\tilde{s}_{\beta\alpha}\right)c_{i\alpha}.
\end{equation}
Subsequently, the quantities with the superscript ``*'' are updated to account for the source terms:
\begin{align}
     & \tilde{p}\left(\bm{x},t\right)=\tilde{p}^*\left(\bm{x},t\right)+\sum_{i=0}^{Q-1}w_i\left\{-3\Delta x\kappa\left(\bm{x},t\right)c_{i\alpha}\tilde{u}_\alpha\left(\bm{x},t\right)-\Delta x\frac{\partial J_\mathcal{O}}{\partial f_i}\left(\bm{x},t\right)\right\}, \label{eq:ap}                                                  \\
     & \tilde{u}_\alpha\left(\bm{x},t\right)=\tilde{u}_\alpha^*\left(\bm{x},t\right)+\sum_{i=0}^{Q-1}w_ic_{i\alpha}\left\{-3\Delta x\kappa\left(\bm{x},t\right)c_{i\beta}\tilde{u}_\beta\left(\bm{x},t\right)-\Delta x\frac{\partial J_\mathcal{O}}{\partial f_i}\left(\bm{x},t\right)\right\},                                         \\
     & \tilde{s}_{\alpha\beta}\left(\bm{x},t\right)=\tilde{s}_{\alpha\beta}^*\left(\bm{x},t\right)+\sum_{i=0}^{Q-1}w_ic_{i\alpha}c_{i\beta}\left\{-3\Delta x\kappa\left(\bm{x},t\right)c_{i\gamma}\tilde{u}_\gamma\left(\bm{x},t\right)-\Delta x\frac{\partial J_\mathcal{O}}{\partial f_i}\left(\bm{x},t\right)\right\}. \label{eq:as}
\end{align}

On the other hands, $\tilde{u}_\alpha^\text{G}$ and $\tilde{\omega}$ are given by
\begin{align}
     & \tilde{u}_\alpha^\text{G}\left(t\right)=-\int_\mathcal{O}\int_D3\kappa\left(\bm{x},t\right)\tilde{u}_\alpha\left(\bm{x},t\right)W\left(\bm{x},\bm{x}^\text{ref}\left(\bm{\xi},t\right)\right)d\Omega^\text{ref}d\Omega+\frac{\partial J_\mathcal{I}}{\partial u_\alpha^\text{G}}\left(t\right),                                                      \\
     & \tilde{\omega}\left(t\right)=-\int_\mathcal{O}\int_D3\kappa\left(\bm{x},t\right)\tilde{u}_\alpha\left(\bm{x},t\right)\frac{\partial u_\alpha^\text{S,ref}}{\partial\omega}\left(\bm{\xi},t\right)W\left(\bm{x},\bm{x}^\text{ref}\left(\bm{\xi},t\right)\right)d\Omega^\text{ref}d\Omega+\frac{\partial J_\mathcal{I}}{\partial\omega}\left(t\right).
\end{align}
From Eq.~\eqref{eq:rigid_body_velocity}, $\partial u_\alpha^\text{S,ref}/\partial\omega$ is expressed as
\begin{equation}
    \left(\begin{array}{c}\displaystyle\frac{\partial u_1^\text{S,ref}}{\partial\omega}\vspace{2pt} \\ \displaystyle\frac{\partial u_2^\text{S,ref}}{\partial\omega}\end{array}\right)=\left(\begin{array}{cc}-\sin\theta\left(t\right) & -\cos\theta\left(t\right) \\
             \cos\theta\left(t\right)      & -\sin\theta\left(t\right)\end{array}\right)\left(\begin{array}{c}\xi_1-\xi_1^\text{G}\\\xi_2-\xi_2^\text{G}\end{array}\right).
\end{equation}

\subsection{Objective and constraint functionals}
\label{sec26}

We consider two types of objective functionals.
The first functional represents a weighted combination of the mean and variance of the volumetric flow rate in a prescribed direction, whereas the second corresponds to a weighted measure of the propulsive performance and hydrodynamic resistance acting on a rigid body, such as an oar during a rowing motion.
Here, the first objective functional is employed in Section~\ref{sec41} and \ref{sec43}, while the second one is employed in Section~\ref{sec42}.
The first objective functional is defined as
\begin{align}
     & J_1=E-\phi V, \label{eq:objective1}                                                                                                                                                                                                                                                                                             \\
     & E=\frac{\int_\mathcal{I}\int_{\mathcal{O}_\text{target}}n_\alpha u_\alpha d\Omega dt}{\int_\mathcal{I}\int_{\mathcal{O}_\text{target}}d\Omega dt},                                                                                                                                                                              \\
     & V=\frac{\int_\mathcal{I}\int_{\mathcal{O}_\text{target}}\left(n_\alpha u_\alpha\right)^2d\Omega dt}{\int_\mathcal{I}\int_{\mathcal{O}_\text{target}}d\Omega dt}-\left(\frac{\int_\mathcal{I}\int_{\mathcal{O}_\text{target}}n_\alpha u_\alpha d\Omega dt}{\int_\mathcal{I}\int_{\mathcal{O}_\text{target}}d\Omega dt}\right)^2,
\end{align}
where the first term on the right hand side, $E$, represents the temporal and spatial average of the flow velocity component in the prescribed direction, and the second term, $V$, corresponds to its variance.
The parameter $\phi$ is a weighting factor, the specific value of which is given in Section~\ref{sec41} and \ref{sec43}.
Here, $\mathcal{O}_\text{target}$ denotes the target region over which the flow rate is evaluated, and $n_\alpha$ specifies the direction along which the flow rate is measured.

The second objective functional is defined as
\begin{align}
     & J_2=F_L+w_DF_D+w_TT, \label{eq:objective2}                                                                                                                                                                     \\
     & F_L=\frac{\int_\mathcal{I}\int_D\kappa^\text{ref}\left(u_1^\text{ref}-u_1^\text{S,ref}\right)d\Omega dt}{\int_\mathcal{I}\int_Dd\Omega dt},                                                                    \\
     & F_D=\frac{\int_\mathcal{I}u_2^\text{G}\int_D\kappa^\text{ref}\left(u_2^\text{ref}-u_2^\text{S,ref}\right)d\Omega dt}{\left(x_2^\text{max}-x_2^\text{min}\right)\int_Dd\Omega},                                 \\
     & T=\frac{\int_\mathcal{I}\omega\int_De_{3\alpha\beta}\left(x_\alpha^\text{ref}-x_\alpha^\text{G}\right)\kappa^\text{ref}\left(u_\beta^\text{ref}-u_\beta^\text{S,ref}\right)d\Omega dt}{2\pi l_m\int_Dd\Omega},
\end{align}
where the first term on the right hand side, $F_L$, represents the propulsive force, the second term, $F_D$, represents the work done by the drag force, and the third term, $T$, represents the work done by the resistive torque, respectively.
The parameter $w_D$ and $w_T$ is weighting factors for $F_D$ and $T$, respectively, and the specific values of which are given in Section~\ref{sec42}.
$l_m$ denotes the characteristic moment arm length, and its specific value is also given in Section~\ref{sec42}.

Contrary, we consider three types of constraint functionals.
The first functional represents a solid volume constraint, whereas the second corresponds to a fluid volume constraint, and the third functional represents the maximum angular velocity.
The first constraint functional is employed in Section~\ref{sec41}, whereas the second and third are both employed in Section~\ref{sec42}.
The first and second constraint functionals are defined as
\begin{align}
     & G_1=\frac{\int_D\gamma d\Omega}{V_\text{max}\int_Dd\Omega}-1, \label{eq:constraint1}                \\
     & G_2=\frac{\int_D\left(1-\gamma\right) d\Omega}{V_\text{max}\int_Dd\Omega}-1, \label{eq:constraint2}
\end{align}
where $V_\text{max}$ denotes the maximum allowable volume ratio.
The specific value is provided in Section~\ref{sec41} and \ref{sec42}.

The third constraint functional is defined as
\begin{equation}
    G_3=\left\{\int_\mathcal{I}\left(\frac{\omega}{\omega_\text{max}}\right)^pdt\right\}^\frac{1}{p}-1, \label{eq:constraint3}
\end{equation}
where $\omega_\text{max}$ is the prescribed upper bound of the angular velocity.
This constraint is formulated using the p-norm of the angular velocity, which enables the maximum value to be controlled through a single functional.
The parameter $p$ is chosen as a sufficiently large positive value.
The p-norm formulation is widely used to approximate maximum-type constraints, such as maximum stress constraints in structural optimization~\cite{https://doi.org/10.1002/nme.6781}.
The specific values of $\omega_\text{max}$ and $p$ are also provided in Section~\ref{sec42}.

\section{Numerical implementation}
\label{sec3}

\subsection{Filtering scheme}
\label{sec31}

We employ a density filter with a Heaviside projection~\cite{wang2011projection} to simplify the obtained shape and to promote a clear binarized design.
First, the design variable $\gamma$, defined at each grid point, is filtered as follows:
\begin{equation}
    \tilde{\gamma}_i=\frac{\sum_jw_\text{s}\left(\bm{x}_i,\bm{x}_j\right)\gamma_j}{\sum_jw_\text{s}\left(\bm{x}_i,\bm{x}_j\right)}, \label{eq:filtering}
\end{equation}
where quantities with subscripts $i$ and $j$ are defined at the $i$-th and $j$-th grid points, respectively.
The weight function $w_\text{s}$ is given by
\begin{equation}
    w_\text{s}\left(\bm{x}_i,\bm{x}_j\right)=\left\{\begin{array}{ll}\frac{R-\left\|\bm{x}_i-\bm{x}_J\right\|}{R}&\left\|\bm{x}_i-\bm{x}_j\right\|\leq R\vspace{2pt}\\0&\left\|\bm{x}_i-\bm{x}_j\right\|\geq R\end{array}\right.,
\end{equation}
where $R$ is the filter radius, and its specific value is given in Section~\ref{sec4}.
The filtered design variable $\tilde{\gamma}$ is projected as follows:
\begin{equation}
    \bar{\tilde{\gamma}}_i=\frac{\tanh{\left(\beta\eta\right)}+\tanh{\left(\beta\left(\tilde{\gamma}_i-\eta\right)\right)}}{\tanh{\left(\beta\eta\right)}+\tanh{\left(\beta\left(1-\eta\right)\right)}}. \label{eq:projection}
\end{equation}
Here, $\beta$ and $\eta$ denotes the steepness parameter and the threshold value, respectively.
The specific values of $\beta$ and $\eta$ are also provided in Section~\ref{sec4}.

\subsection{Continuation scheme}
\label{sec32}

We employ a continuation scheme~\cite{Stolpe2001} to stabilize the optimization steps.
In this scheme, the parameters are updated throughout the optimization process.
In the present study, the updated parameters are the steepness parameter $\beta$ in the Heaviside projection defined in Eq.~\eqref{eq:projection}.

First, the parameter is fixed at a prescribed value, and the optimization iterations are performed until either
(i) the prescribed number of iterations, $N_\text{opt}$, is reached, or
(ii) the stopping criterion is satisfied and all the constraint functionals are satisfied.
Next, the parameter is updated, and the optimization iterations are performed again.
This parameter update procedure is repeated $N_\text{cont}$ times, where $N_\text{cont}$ denotes the prescribed number of continuation steps.

The stopping criterion is defined as follows:
\begin{equation}
    \frac{\left|J^{k+1}-J^k\right|}{\left|J^k\right|}\leq 1\times10^{-6},
\end{equation}
where $J^{k}$ and $J^{k+1}$ denote the objective functional values at the $k$-th and $k+1$-th optimization step.

\subsection{Periodic problem approximation}
\label{sec33}

We employ a periodic problem approximation~\cite{Tanabe2023} to reduce computational cost.
In this method, the terminal values of the state fields at the previous optimization step are reused as the initial values at the current optimization step.
In other words, for example:
\begin{equation}
    p^{k}\left(t_0\right)=p^{k-1}\left(t_1\right),
\end{equation}
where the superscripts denote the optimization step number.
The adjoint fields are treated in the same way; the initial values at the previous step are reused as the terminal values, for example:
\begin{equation}
    \tilde{p}^{k}\left(t_1\right)=\tilde{p}^{k-1}\left(t_0\right).
\end{equation}
Here, the adjoint fields are computed in reverse time; therefore, the reused time is switched compared with the state fields.

\subsection{Optimization procedure}
\label{sec34}

The optimization procedure is shown in Algorithm~\ref{algorithm:optimization_procedure}.
Here, we employ the Method of Moving Asymptotes (MMA)~\cite{svanberg1987method}, a gradient-based optimizer.

\begin{algorithm}
    \small

    \SetKw{KwInit}{Initialize}

    \caption{Optimization procedure}

    \KwInit{$\gamma,\chi_\alpha,\psi$}\;
    \For{$l=1$ to $N_\text{cont}$}{
        $\beta=\beta^\text{l}$\;
        \For{$k=1$ to $N_\text{opt}$}{
            \tcp{Filter variables and inject design variable into governing equations}
            $\gamma\mapsto\tilde{\gamma},\tilde{\gamma},\beta\mapsto\bar{\tilde{\gamma}},\bar{\tilde{\gamma}}\mapsto\kappa^\text{ref}$\tcp*{Eqs.~\eqref{eq:filtering}, \eqref{eq:projection} and \eqref{eq:kappa_ref}}
            $\chi\mapsto\tilde{\chi},\tilde{\chi}\mapsto x_\alpha^\text{G},x_\alpha^\text{G}\mapsto u_\alpha^\text{G}$\tcp*{Eqs.~\eqref{eq:chi_filtered}, \eqref{eq:x_g} and \eqref{eq:u_g}}
            $\psi\mapsto\tilde{\psi},\tilde{\psi}\mapsto\theta,\theta\mapsto\omega$\tcp*{Eqs.~\eqref{eq:psi_filtered}, \eqref{eq:theta} and \eqref{eq:omega}}
            \BlankLine
            \tcp{Compute state fields}
            \KwInit{$p\left(t_0\right)$, $u_\alpha\left(t_0\right)$}\;
            \For{$t=1$ to $N_\text{t}$}{
                $p\left(t-\Delta t\right),u_\alpha\left(t-\Delta t\right)\mapsto p^*\left(t\right),u_\alpha^*\left(t\right)$\tcp*{Eqs.~\eqref{eq:p_tmp} and \eqref{eq:u_tmp}}\
                $x^\text{G}_\alpha\left(t\right),u^\text{G}_\alpha\left(t\right),\theta\left(t\right),\omega\left(t\right)\mapsto x^\text{ref}_\alpha\left(t\right),u^\text{S,ref}_\alpha\left(t\right)$\tcp*{Eqs.~\eqref{eq:rigid_body_position} and \eqref{eq:rigid_body_velocity}}\
                $\kappa^\text{ref},x^\text{ref}_\alpha\left(t\right),u^\text{S,ref}_\alpha\left(t\right)\mapsto\kappa\left(t\right),u^\text{S}_\alpha\left(t\right)$\tcp*{Eqs.~\eqref{eq:kappa} and \eqref{eq:u_s}}\
                $p^*\left(t\right),u^*_\alpha\left(t\right),\kappa,u^\text{S}_\alpha\left(t\right)\mapsto p\left(t\right),u_\alpha\left(t\right)$\tcp*{Eqs.~\eqref{eq:p} and \eqref{eq:u}}
            }
            \BlankLine
            \tcp{Compute objective and constraint and converse check}
            $p,u_\alpha\mapsto J,G_j$\;
            \If{Is stopping criterion satisfied}{
                Bleak\;
            }
            \BlankLine
            \tcp{Compute adjoint fields}
            \KwInit{$\tilde{p}\left(t_1\right)$, $\tilde{u}_\alpha\left(t_1\right)$, $\tilde{s}_{\alpha\beta}\left(t_1\right)$}\;
            \For{$t=1$ to $N_\text{t}$}{
                $\tilde{p}\left(t+\Delta t\right),\tilde{u}_\alpha\left(t+\Delta t\right),\tilde{s}_{\alpha\beta}\left(t+\Delta t\right)\mapsto\tilde{p}^*\left(t\right),\tilde{u}^*_\alpha\left(t\right),\tilde{s}^*_{\alpha\beta}\left(t\right)$\tcp*{Eqs.~\eqref{eq:ap_tmp}-\eqref{eq:as_tmp}}\
                $x^\text{G}_\alpha\left(t\right),u^\text{G}_\alpha\left(t\right),\theta\left(t\right),\omega\left(t\right)\mapsto x^\text{ref}_\alpha\left(t\right),u^\text{S,ref}_\alpha\left(t\right)$\tcp*{Eqs.~\eqref{eq:rigid_body_position} and \eqref{eq:rigid_body_velocity}}\
                $\kappa^\text{ref},x^\text{ref}_\alpha\left(t\right),u^\text{S,ref}_\alpha\left(t\right)\mapsto\kappa\left(t\right),u^\text{S}_\alpha\left(t\right)$\tcp*{Eqs.~\eqref{eq:kappa} and \eqref{eq:u_s}}\
                $\tilde{p}^*\left(t\right),\tilde{u}^*_\alpha\left(t\right),\tilde{s}^*_{\alpha\beta}\left(t\right),\kappa\left(t\right),u^\text{S}_\alpha\left(t\right)\mapsto\tilde{p}\left(t\right),\tilde{u}_\alpha\left(t\right),\tilde{s}_{\alpha\beta}\left(t\right)$\tcp*{Eqs.~\eqref{eq:ap}-\eqref{eq:as}}
            }
            \BlankLine
            \tcp{Compute design sensitivity}
            $u_\alpha,\tilde{u}_\alpha,x^\text{G}_\alpha,u^\text{G}_\alpha,\theta,\omega\mapsto\delta J/\delta\bar{\tilde{\gamma}},\delta G_j/\delta\bar{\tilde{\gamma}},\delta J/\delta\tilde{\chi}_\alpha,\delta G_j/\delta\tilde{\chi}_\alpha,\delta J/\delta\tilde{\psi},\delta G_j/\delta\tilde{\psi}$\tcp*{Eqs.~\eqref{eq:dJdgamma},\eqref{eq:dJdchi} and \eqref{eq:dJdpsi}}
            \BlankLine
            \tcp{Unfilter design sensitivity}
            $\delta J/\delta\bar{\tilde{\gamma}}\mapsto\delta J/\delta\gamma,\delta G_j/\delta\bar{\tilde{\gamma}}\mapsto\delta G_j/\delta\gamma$\;
            $\delta J/\delta\tilde{\chi}_\alpha\mapsto\delta J/\delta\chi_\alpha,\delta G_j/\delta\tilde{\chi}_\alpha\mapsto\delta G_j/\delta\chi_\alpha$\;
            $\delta J/\delta\tilde{\psi}\mapsto\delta J/\delta\psi,\delta G_j/\delta\tilde{\psi}\mapsto\delta G_j/\delta\psi$\tcp*{Chain rule}
            \BlankLine
            \tcp{Update design variables}
            $\gamma^k,\chi_\alpha^k,\psi^k,\delta J/\delta\gamma,\delta J/\delta\chi_\alpha,\delta J/\delta\psi,G_j,\delta G_j/\delta\gamma,\delta G_j/\delta\chi_\alpha,\delta G_j/\delta\psi\mapsto\gamma^{k+1},\chi_\alpha^{k+1},\psi^{k+1}$\tcp*{MMA}
        }
    }

    \label{algorithm:optimization_procedure}
\end{algorithm} 

\section{Numerical examples}
\label{sec4}

We have applied the proposed method for three kinds of problems involving moving rigid bodies in fluid flow.
In this section, we will discuss two two-dimensional examples and one three-dimensional example.
All examples are implemented using in-house CUDA/C++ code.
The two-dimensional examples are computed on either a single NVIDIA GeForce RTX 3050 6GB Laptop GPU or a single NVIDIA GeForce RTX 4070 GPU, whereas the three-dimensional example is computed on a single NVIDIA GeForce RTX 4070 GPU.

\subsection{2D pump}
\label{sec41}

\begin{figure}[t]
    \centering
    \begin{subfigure}[b]{0.33\columnwidth}
        \centering
        \includegraphics[width=0.8\columnwidth]{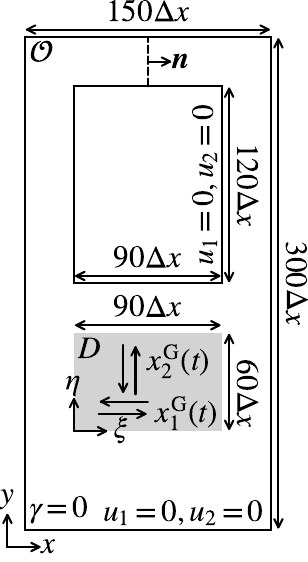}
        \caption{Design setting}
        \label{fig:example1_design_setting}
    \end{subfigure}
    \begin{subfigure}[b]{0.33\columnwidth}
        \centering

        \vspace{0pt}
        \includegraphics[width=0.7\columnwidth]{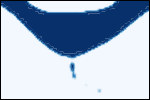}
        \caption{Optimized shape}
        \label{fig:example1_optimized_shape}

        \vspace{0.5em}
        \includegraphics[width=0.7\columnwidth]{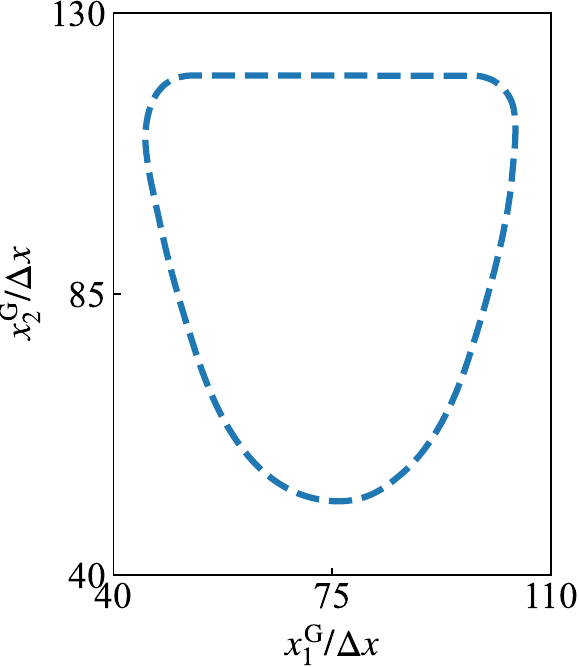}
        \caption{Optimized trajectory}
        \label{fig:example1_optimized_trajectory}
    \end{subfigure}
    \begin{subfigure}[b]{0.33\columnwidth}
        \centering
        \includegraphics[width=\columnwidth]{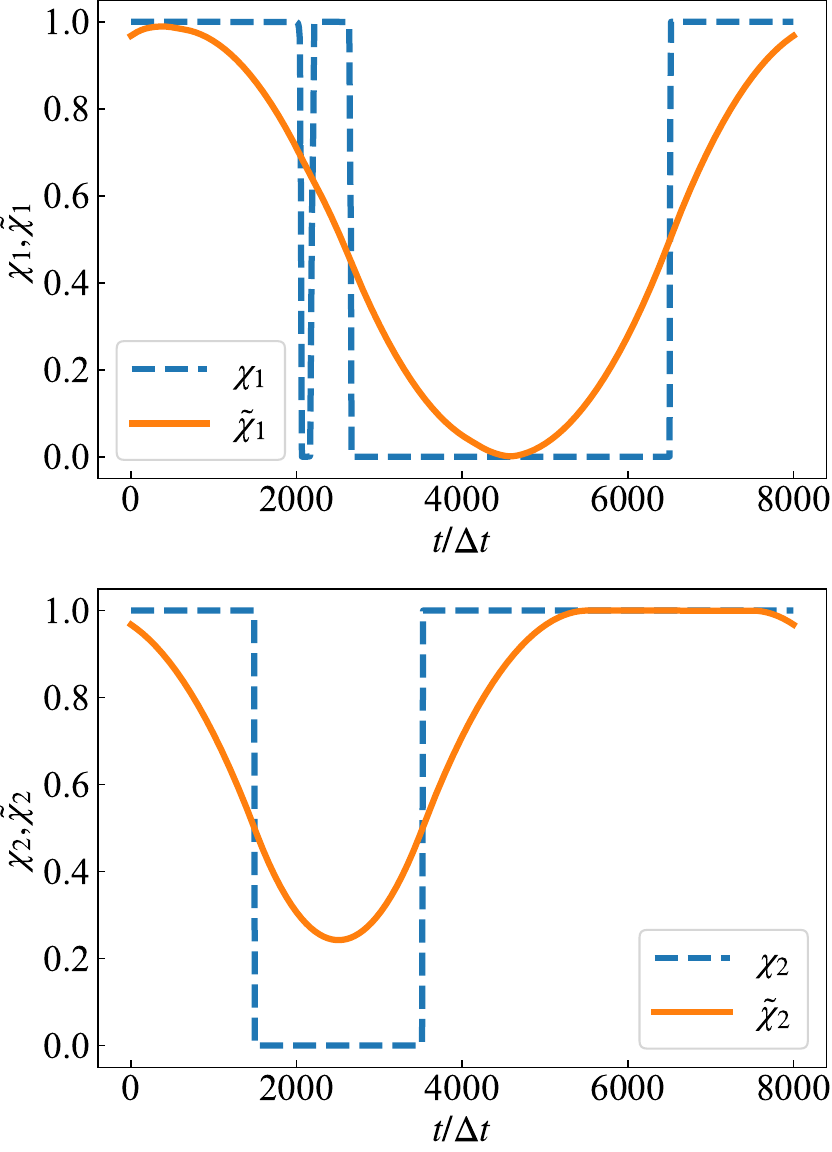}
        \caption{Optimized motion}
        \label{fig:example1_optimized_motion}
    \end{subfigure}
    \caption{Design setting, optimized shape, optimized trajectory, and optimized motion for the \textit{2D pump}.}
\end{figure}
\begin{figure}[t]
    \centering
    \begin{minipage}[t]{0.195\columnwidth}
        \centering
        \includegraphics[width=0.9\columnwidth]{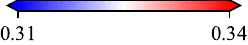}
    \end{minipage}
    \begin{minipage}[t]{0.195\columnwidth}
        \centering
        \includegraphics[width=0.9\columnwidth]{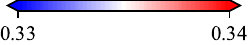}
    \end{minipage}
    \begin{minipage}[t]{0.195\columnwidth}
        \centering
        \includegraphics[width=0.9\columnwidth]{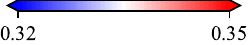}
    \end{minipage}
    \begin{minipage}[t]{0.195\columnwidth}
        \centering
        \includegraphics[width=0.9\columnwidth]{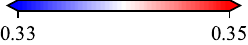}
    \end{minipage}
    \begin{minipage}[t]{0.195\columnwidth}
        \centering
        \includegraphics[width=0.9\columnwidth]{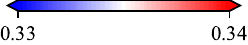}
    \end{minipage} \\
    \begin{minipage}[t]{0.195\columnwidth}
        \centering
        \includegraphics[width=0.9\columnwidth]{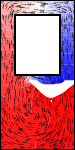}
        \subcaption{$t/\Delta t=500$}
    \end{minipage}
    \begin{minipage}[t]{0.195\columnwidth}
        \centering
        \includegraphics[width=0.9\columnwidth]{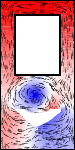}
        \subcaption{$t/\Delta t=2500$}
    \end{minipage}
    \begin{minipage}[t]{0.195\columnwidth}
        \centering
        \includegraphics[width=0.9\columnwidth]{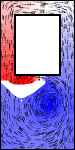}
        \subcaption{$t/\Delta t=4500$}
    \end{minipage}
    \begin{minipage}[t]{0.195\columnwidth}
        \centering
        \includegraphics[width=0.9\columnwidth]{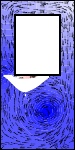}
        \subcaption{$t/\Delta t=5500$}
    \end{minipage}
    \begin{minipage}[t]{0.195\columnwidth}
        \centering
        \includegraphics[width=0.9\columnwidth]{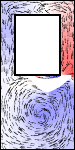}
        \subcaption{$t/\Delta t=7500$}
    \end{minipage}
    \caption{Pressure distributions and normalized flow velocity vector plots at each time step for the \textit{2D pump}.}
    \label{fig:example1_pressure}
\end{figure}
\begin{figure}
    \centering
    \includegraphics[width=0.5\columnwidth]{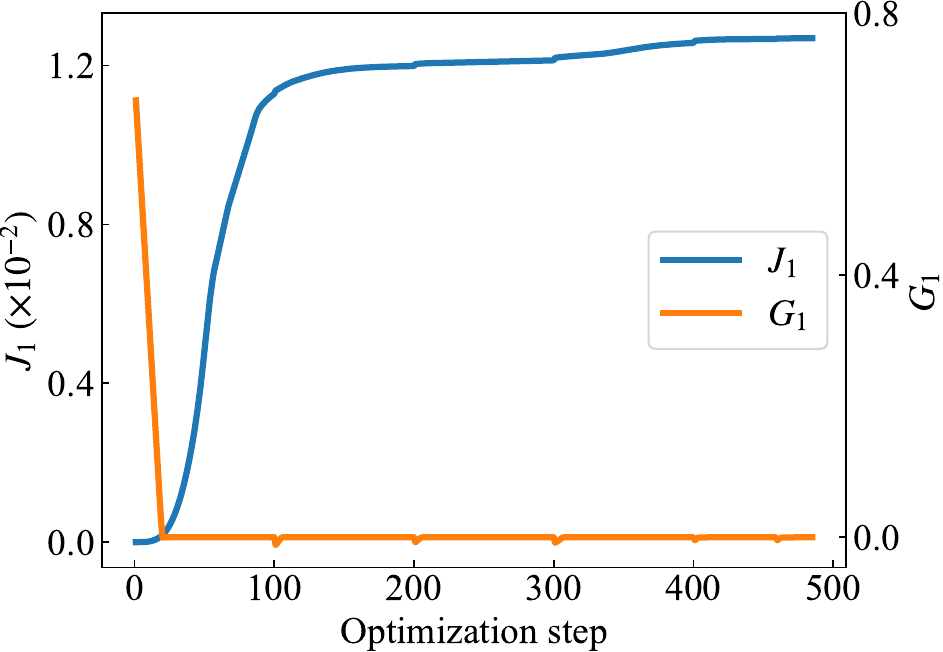}
    \caption{Histories of the objective and constraint functional values for the \textit{2D pump}.}
    \label{fig:example1_history}
\end{figure}
\begin{figure}[t]
    \centering
    \begin{minipage}[t]{0.49\columnwidth}
        \centering
        \includegraphics[width=0.8\columnwidth]{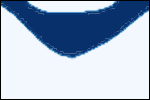}
        \subcaption{Optimized shape without islands}
        \label{fig:example1_optimized_shape_mod}
    \end{minipage}
    \begin{minipage}[t]{0.49\columnwidth}
        \centering
        \includegraphics[width=0.8\columnwidth]{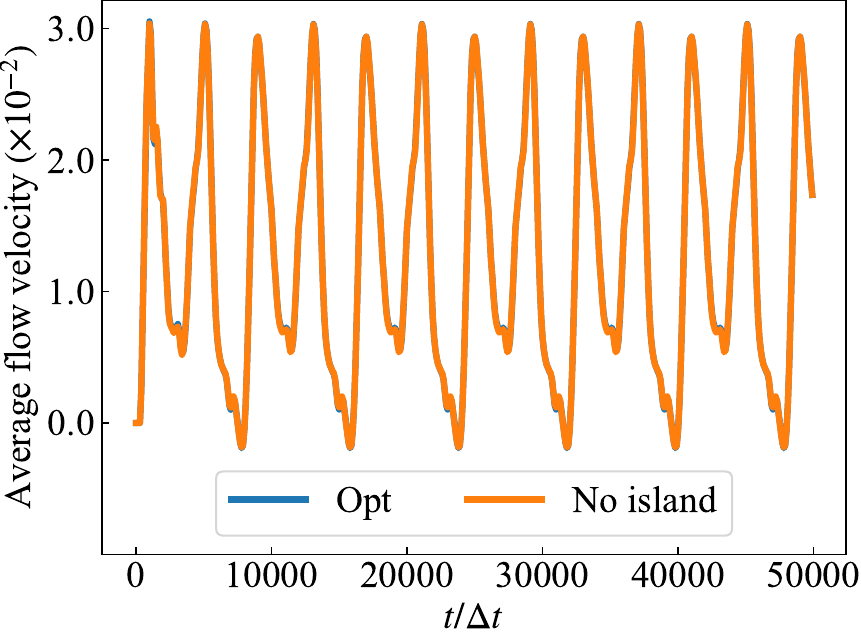}
        \subcaption{Average flow velocity}
        \label{fig:example1_compare_island}
    \end{minipage}
    \caption{Optimized shape without islands and average flow velocity at each time step for the original optimized shape and the modified optimized shape without islands.}
\end{figure}
\begin{figure}[t]
    \centering
    \begin{subfigure}[b]{0.33\columnwidth}
        \centering

        \vspace{0pt}
        \includegraphics[width=0.7\columnwidth]{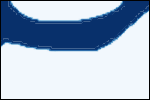}
        \caption{Optimized shape obtained by shape-only optimization}
        \label{fig:example1_optimized_shape_so}

        \vspace{0.5em}
        \includegraphics[width=0.7\columnwidth]{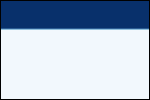}
        \caption{Prescribed shape used in motion-only optimization}
        \label{fig:example1_optimized_shape_mo}
    \end{subfigure}
    \begin{subfigure}[b]{0.33\columnwidth}
        \centering
        \includegraphics[width=0.9\columnwidth]{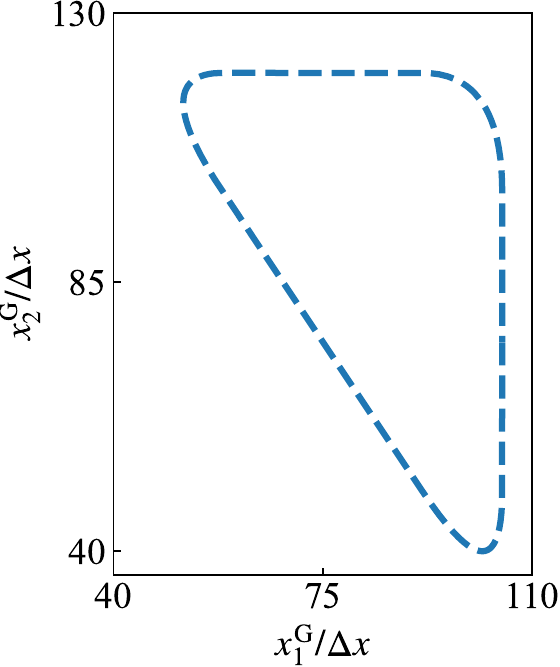}
        \caption{Prescribed trajectory used in shape-only optimization}
        \label{fig:example1_optimized_trajectory_so}
    \end{subfigure}
    \begin{subfigure}[b]{0.33\columnwidth}
        \centering
        \includegraphics[width=0.9\columnwidth]{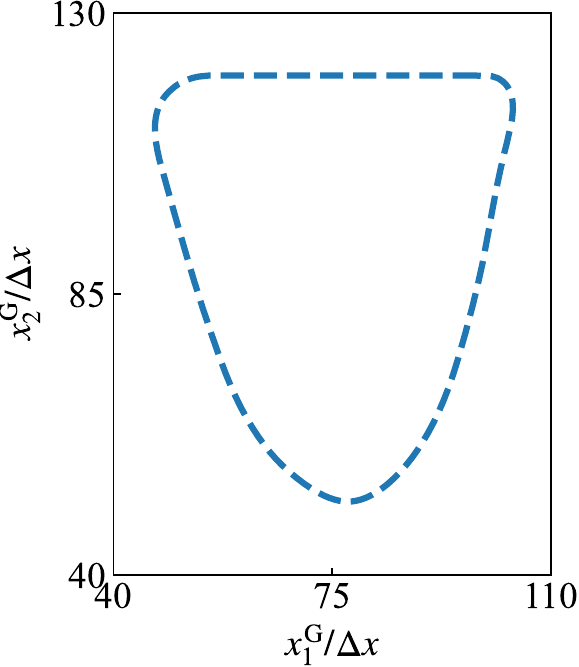}
        \caption{Optimized trajectory obtained by motion-only optimization}
        \label{fig:example1_optimized_trajectory_mo}
    \end{subfigure}
    \caption{Optimized or prescribed shapes and trajectories for the shape-only and motion-only optimization cases in the \textit{2D pump}.}
\end{figure}
\begin{table}[t]
    \centering
    \caption{Objective functional values for each optimization case of the \textit{2D pump}.}
    \begin{tabular}{|c|c|c|} \hline
        Shape and motion & Shape only & Motion only \\ \hline
        0.0126878        & 0.0104377  & 0.0113832   \\ \hline
    \end{tabular}
    \label{table:example1_objective_comparing_optimization}
\end{table}
\begin{figure}[t]
    \centering
    \begin{minipage}[t]{0.33\columnwidth}
        \centering
        \includegraphics[width=\columnwidth]{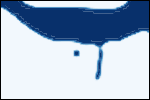}
        \subcaption{$N_\text{t}=6000\Delta t$}
    \end{minipage}
    \begin{minipage}[t]{0.33\columnwidth}
        \centering
        \includegraphics[width=\columnwidth]{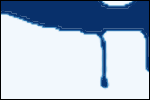}
        \subcaption{$N_\text{t}=10000\Delta t$}
    \end{minipage}
    \begin{minipage}[t]{0.33\columnwidth}
        \centering
        \includegraphics[width=\columnwidth]{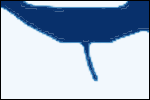}
        \subcaption{$N_\text{t}=12000\Delta t$}
    \end{minipage}
    \caption{Optimized shapes for different values of $N_\text{t}$ in the \textit{2D pump}.}
    \label{fig:example1_optimized_shapes_for_each_nt}
\end{figure}
\begin{figure}[t]
    \centering
    \begin{minipage}[t]{0.33\columnwidth}
        \centering
        \includegraphics[width=\columnwidth]{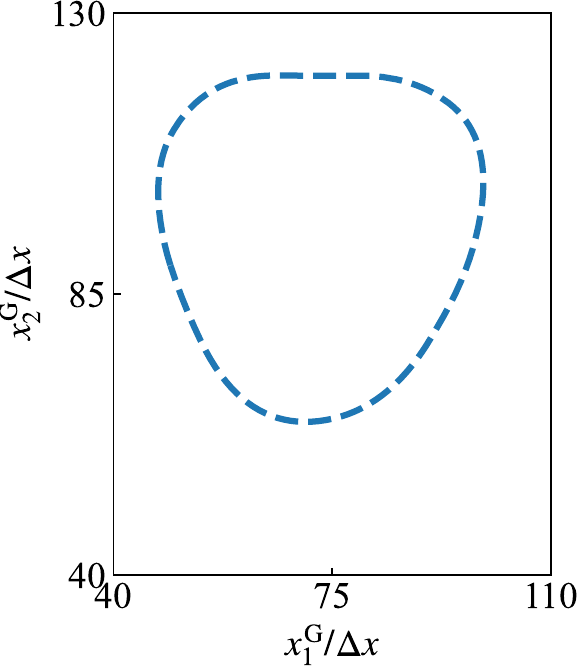}
        \subcaption{$N_\text{t}=6000\Delta t$}
    \end{minipage}
    \begin{minipage}[t]{0.33\columnwidth}
        \centering
        \includegraphics[width=\columnwidth]{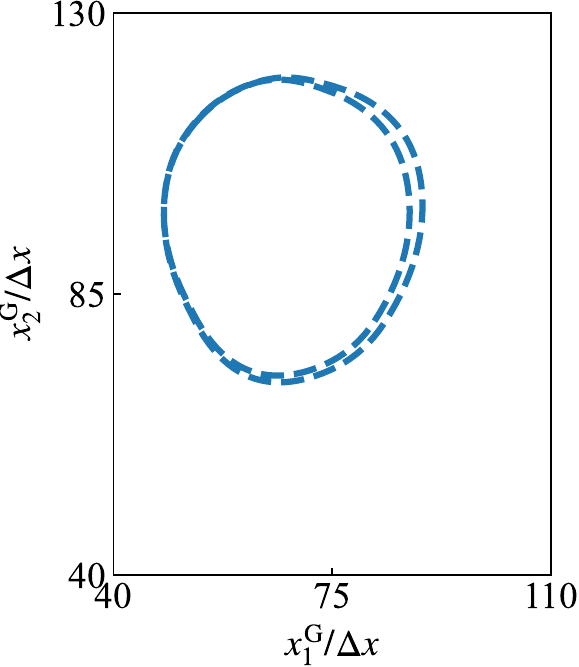}
        \subcaption{$N_\text{t}=10000\Delta t$}
    \end{minipage}
    \begin{minipage}[t]{0.33\columnwidth}
        \centering
        \includegraphics[width=\columnwidth]{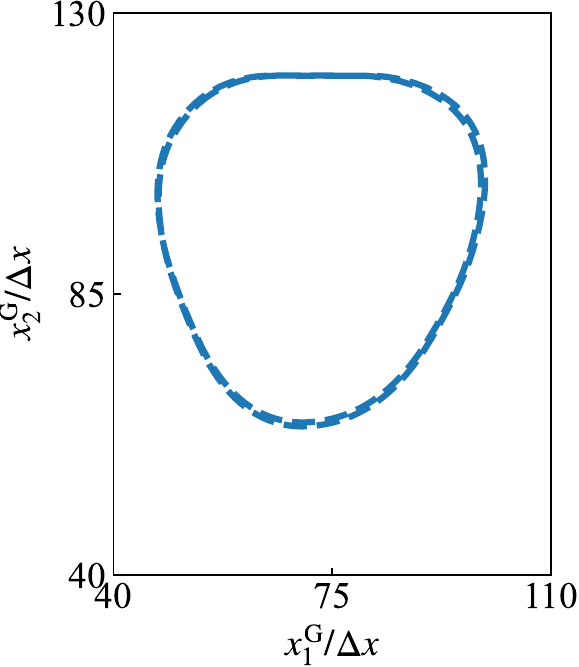}
        \subcaption{$N_\text{t}=12000\Delta t$}
    \end{minipage}
    \caption{Optimized trajectories for different values of $N_\text{t}$ in the \textit{2D pump}.}
    \label{fig:example1_optimized_trajectory_for_each_nt}
\end{figure}
\begin{figure}[t]
    \centering
    \begin{subfigure}[b]{0.33\columnwidth}
        \centering

        \vspace{0pt}
        \includegraphics[width=0.7\columnwidth]{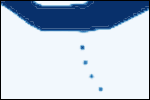}
        \caption{Optimized shape for $V_\text{max}=22.5\%$}

        \vspace{0.5em}
        \includegraphics[width=0.7\columnwidth]{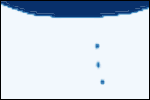}
        \caption{Optimized shape for $V_\text{max}=15\%$}
    \end{subfigure}
    \begin{subfigure}[b]{0.33\columnwidth}
        \centering
        \includegraphics[width=0.9\columnwidth]{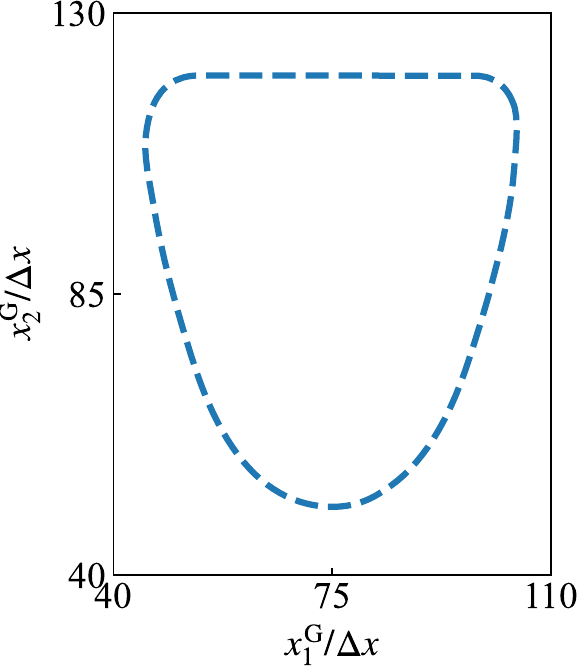}
        \caption{Optimized trajectory for $V_\text{max}=22.5\%$}
    \end{subfigure}
    \begin{subfigure}[b]{0.33\columnwidth}
        \centering
        \includegraphics[width=0.9\columnwidth]{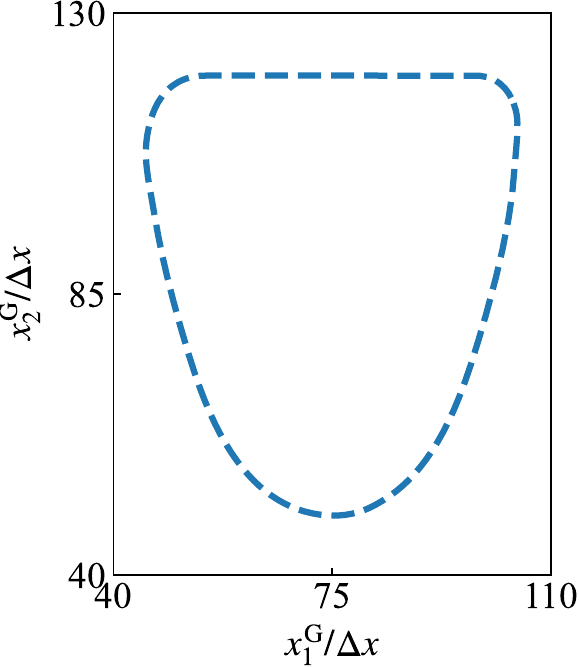}
        \caption{Optimized trajectory for $V_\text{max}=15\%$}
    \end{subfigure}
    \caption{Optimized shapes and trajectories for different values of $V_\text{max}$ in the \textit{2D pump}.}
    \label{fig:example1_optimized_shapes_and_trajectories_for_each_vmax}
\end{figure}

As a first example, we consider the problem of simultaneously designing the shape and translational motion in two dimensions, referred to as a \textit{2D pump}.
As shown in Fig.~\ref{fig:example1_design_setting}, the analysis domain $\mathcal{O}$ is a rectangle with an inner wall, and the design domain $D$ undergoes only translational motion based on the design variable $\chi_\alpha$, where $\alpha$ takes the values $1$ and $2$.
The objective functional is the flow velocity maximization, $J_1$, which is defined in Eq.~\eqref{eq:objective1} with $\phi=0$.
In other words, only the average flow velocity is considered.
$\mathcal{O}_\text{target}$ is indicated by the dashed line, and the direction of the flow to be maximized is denoted by $\bm{n}$, in Fig.~\ref{fig:example1_design_setting}.
The constraint functional is a volume constraint defined in Eq.~\eqref{eq:constraint1}.
The other parameters are as follows: $N_\text{t}=8000\Delta t$, $\nu=0.1\Delta x$, $N_\text{opt}=100$, $\kappa^\text{ref}_\text{max}=300$, $q=0.1$, $\beta=\left\{1,2,4,8,16,32\right\}$, $R=4.8\Delta x$, $R_\text{t}=2000\Delta t$, and $V_\text{max}=0.3$.
Additionally, the periodic approximation described in Section~\ref{sec33} is employed to efficiently solve the periodic problem.
The initial shape is uniformly $\gamma=0.5$, and the initial motion is uniformly $\chi_\alpha=0.5$.
The initial condition of the fluid is $p=1/3$ and $u_\alpha=0$.

The optimized shape is shown in Fig.~\ref{fig:example1_optimized_shape}, while the optimized trajectory is shown in Fig.~\ref{fig:example1_optimized_trajectory}.
In Fig.~\ref{fig:example1_optimized_motion}, the optimized position of the center of the design grid at each time step is shown.
In Fig.~\ref{fig:example1_optimized_motion}, the dashed line represents the design variables without filtering, which are directly optimized by the optimization solver (e.g., MMA), whereas the solid line represents the filtered design variables, i.e., the normalized position within the movable region.
The optimized shape resembles a V-shaped profile, with a cavity at the top and a smooth curve at the bottom.
Additionally, the pressure distribution and normalized flow velocity vector plots at each time step are shown in Fig.~\ref{fig:example1_pressure}.
The optimized motion follows a triangular-like orbit in the lower half of the analysis domain.
During $t=0$ to $t=2500\Delta t$, the rigid body moves from near the right port to the center bottom to prepare for the next motion, which pushes the fluid toward the left port.
At this phase, as shown in $t=500\Delta t$ in Fig.~\ref{fig:example1_pressure}, the pressure on the upper surface of the rigid body with a cavity is relatively low, and the fluid is absorbed.
Between $t=2500\Delta t$ and $t=5000\Delta t$, the rigid body pushes the fluid toward the left port and finally returns to its initial position.
The histories of the objective and constraint functionals are shown in Fig.~\ref{fig:example1_history}.
The objective functional increases monotonically, while the constraint is satisfied at an early stage of the optimization process.
Abrupt changes are observed every $100$ optimization steps.
These changes occur because the steepness parameter $\beta$ is updated at those points according to the continuation scheme described in Section~\ref{sec32}.
There are some islands in the lower half of the optimized shape; therefore, we examine their effect.
We remove the islands a posteriori, as shown in Fig.~\ref{fig:example1_optimized_shape_mod}, and compare the average flow velocity with that of the original optimized shape.
The comparison result is shown in Fig.~\ref{fig:example1_compare_island}.
As shown in the figure, the differences are quite small; in other words, the effects of the islands are negligible.
Therefore, we conclude that they can be removed after the optimization in the practical phase.

We compare the optimized shape and motion with cases of only shape (topology) optimization and only motion optimization.
The optimized shape for the case of only shape optimization is shown in Fig.~\ref{fig:example1_optimized_shape_so}, whereas for the case of only motion optimization, the shape remains the shape shown in Fig.~\ref{fig:example1_optimized_shape_mo} throughout the optimization iterations.
For the case of only shape optimization, the trajectory is defined as shown in Fig.~\ref{fig:example1_optimized_trajectory_so}, and the optimized trajectory for the case of only motion optimization is shown in Fig.~\ref{fig:example1_optimized_trajectory_mo}.
Additionally, the objective functional values for each case, including the case where both shape and motion are optimized, are shown in Table~\ref{table:example1_objective_comparing_optimization}.
As shown in Table~\ref{table:example1_objective_comparing_optimization}, the case where both shape and motion are optimized shows better performance than the cases where only shape or only motion is optimized, indicating the effectiveness of the proposed method.
Focusing on the shape, in both cases, i.e., when optimizing both the shape and motion and when optimizing only the shape, a cavity appears on the top face. This indicates that the cavity on the top face is effective for moving the fluid.
In contrast, focusing on the motion, in both cases, i.e., when optimizing both the shape and motion and when optimizing only the motion, the trajectories are similar to each other. This indicates that the trajectory shown in Fig.~\ref{fig:example1_optimized_trajectory} is suitable for this problem setting.

Next, we examine the effect of the time interval $N_\text{t}$.
The optimized shapes for each $N_\text{t}$ are shown in Fig.~\ref{fig:example1_optimized_shapes_for_each_nt}, and the optimized trajectories for each $N_\text{t}$ are shown in Fig.~\ref{fig:example1_optimized_trajectory_for_each_nt}.
Focusing on the trajectories, when $N_\text{t}$ is small, such as $N_\text{t}=6000\Delta t$ or $N_\text{t}=8000\Delta t$, the trajectories form a single loop, whereas when $N_\text{t}$ is large, such as $N_\text{t}=10000\Delta t$ or $N_\text{t}=12000\Delta t$, the trajectories form double loops.
Interestingly, the shapes and trajectories for $N_\text{t}=6000\Delta t$ and $N_\text{t}=12000\Delta t$ are similar to each other, in the sense that the double-loop trajectory for $N_\text{t}=12000\Delta t$ can be interpreted as two repetitions of the single-loop trajectory for $N_\text{t}=6000\Delta t$.
From these observations, the transition from a single-loop trajectory to a double-loop trajectory is suggested to occur approximately between $N_\text{t}=8000\Delta t$ and $N_\text{t}=10000\Delta t$, and the optimized shape and trajectory appear to exhibit a repetition structure with respect to $N_\text{t}$.

Finally, we examine the effect of the maximum allowable volume ratio $V_\text{max}$ in the volume constraint $G_1$.
The optimized shapes and trajectories for each $V_\text{max}$ are shown in Fig.~\ref{fig:example1_optimized_shapes_and_trajectories_for_each_vmax}.
Focusing on the trajectories, they are the same in all cases, $V_\text{max}=0.15$, $V_\text{max}=0.225$, and $V_\text{max}=0.3$; therefore, a triangle-like trajectory is effective for this problem setting.
In contrast, for the shapes, the cavity on the top face becomes deeper as $V_\text{max}$ increases.
This may be because the cavity on the top face is effective for this problem setting; however, when $V_\text{max}$ is small, the range of representable shapes is limited.

\subsection{2D oar}
\label{sec42}

\begin{figure}[t]
    \centering
    \begin{subfigure}[b]{0.54\columnwidth}
        \centering

        \vspace{0pt}
        \includegraphics[width=0.8\columnwidth]{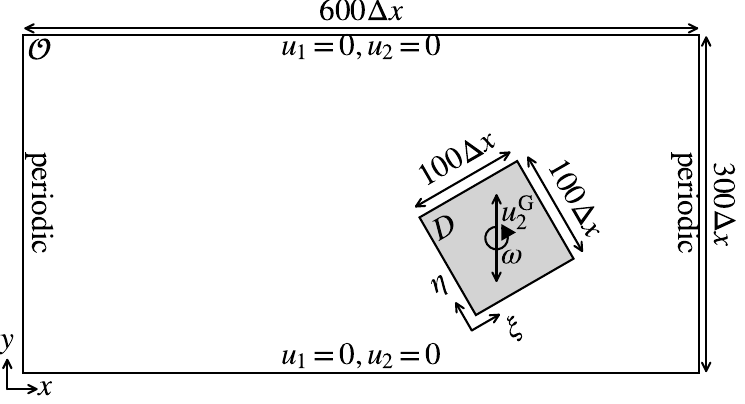}
        \subcaption{Design setting}
        \label{fig:example2_design_setting}

        \vspace{0.5em}
        \includegraphics[width=0.6\columnwidth]{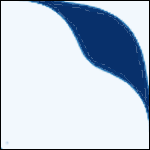}
        \subcaption{Optimized shape}
        \label{fig:example2_optimized_shape}
    \end{subfigure}
    \begin{subfigure}[b]{0.44\columnwidth}
        \centering
        \includegraphics[width=0.9\columnwidth]{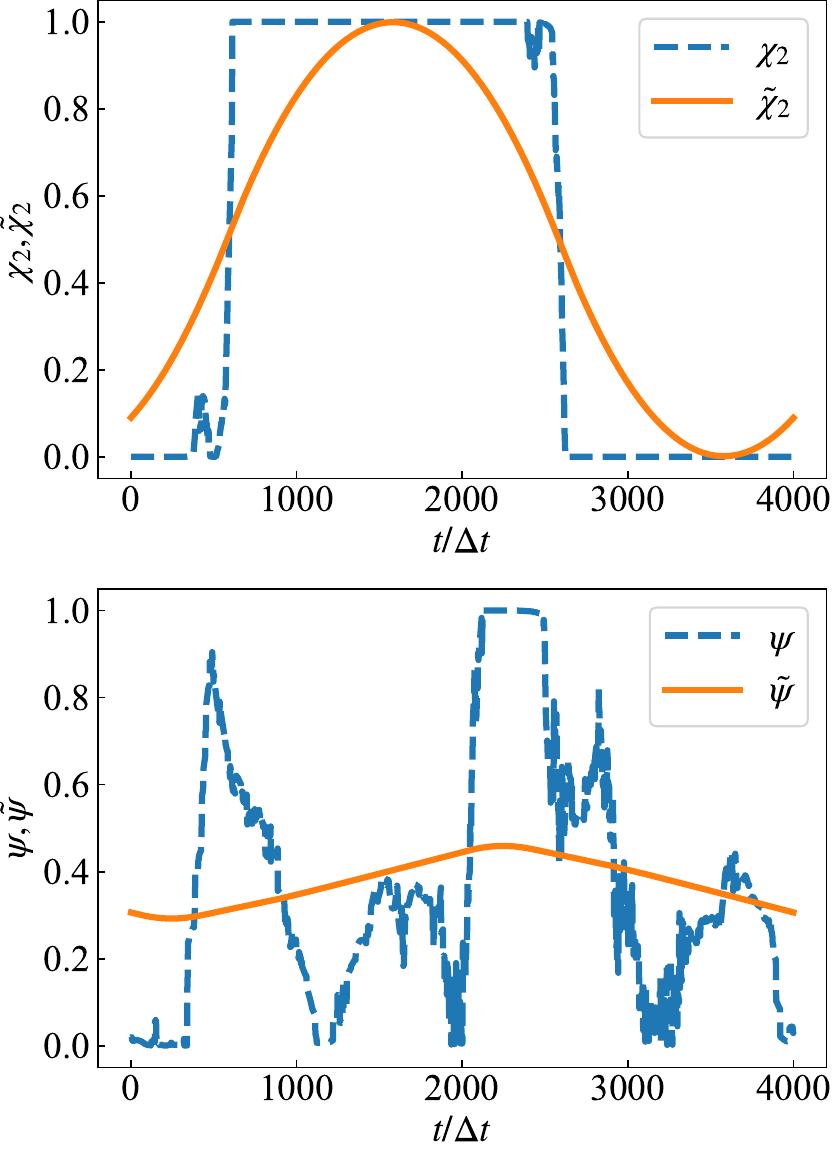}
        \subcaption{Optimized motion}
        \label{fig:example2_optimized_motion}
    \end{subfigure}
    \caption{Design setting, optimized shape and optimized motion for the \textit{2D oar}.}
\end{figure}
\begin{figure}[t]
    \centering
    \begin{minipage}[t]{0.49\columnwidth}
        \centering
        \includegraphics[width=0.9\columnwidth]{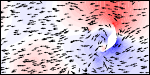}
    \end{minipage}
    \begin{minipage}[t]{0.49\columnwidth}
        \centering
        \includegraphics[width=0.9\columnwidth]{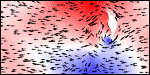}
    \end{minipage} \\
    \begin{minipage}[t]{0.49\columnwidth}
        \centering
        \includegraphics[width=0.4\columnwidth]{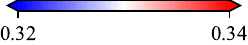}
        \subcaption{$t/\Delta t=600$}
    \end{minipage}
    \begin{minipage}[t]{0.49\columnwidth}
        \centering
        \includegraphics[width=0.4\columnwidth]{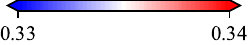}
        \subcaption{$t/\Delta t=1600$}
    \end{minipage} \\
    \begin{minipage}[t]{0.49\columnwidth}
        \centering
        \includegraphics[width=0.9\columnwidth]{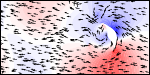}
    \end{minipage}
    \begin{minipage}[t]{0.49\columnwidth}
        \centering
        \includegraphics[width=0.9\columnwidth]{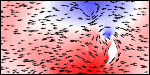}
    \end{minipage} \\
    \begin{minipage}[t]{0.49\columnwidth}
        \centering
        \includegraphics[width=0.4\columnwidth]{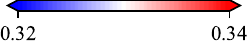}
        \subcaption{$t/\Delta t=2600$}
    \end{minipage}
    \begin{minipage}[t]{0.49\columnwidth}
        \centering
        \includegraphics[width=0.4\columnwidth]{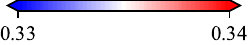}
        \subcaption{$t/\Delta t=3600$}
    \end{minipage}
    \caption{Pressure distributions and normalized velocity vector plots at each time step for the \textit{2D oar}.}
    \label{fig:example2_pressure_at_each_time}
\end{figure}
\begin{figure}
    \centering
    \begin{minipage}[t]{0.49\columnwidth}
        \centering
        \includegraphics[width=0.9\columnwidth]{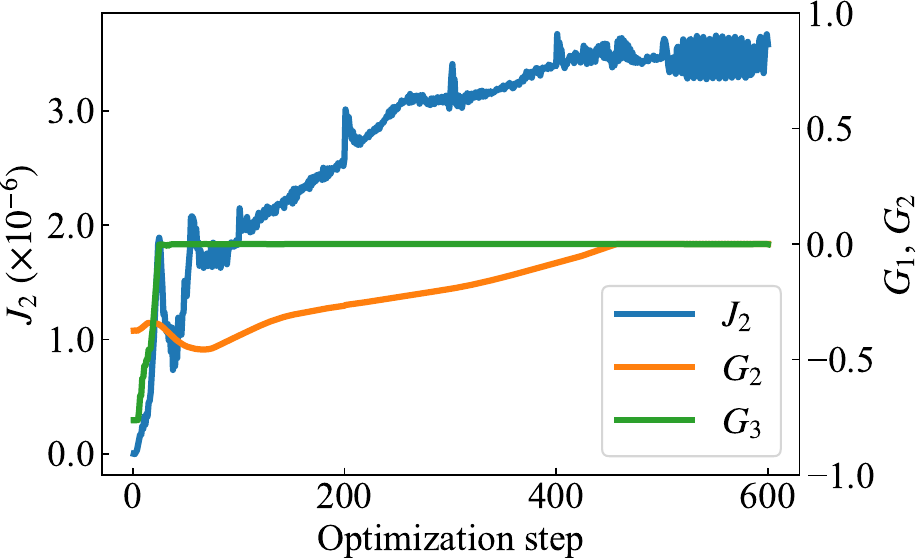}
        \subcaption{History of the objective and constraint functionals}
        \label{fig:example2_history}
    \end{minipage}
    \begin{minipage}[t]{0.49\columnwidth}
        \centering
        \includegraphics[width=0.8\columnwidth]{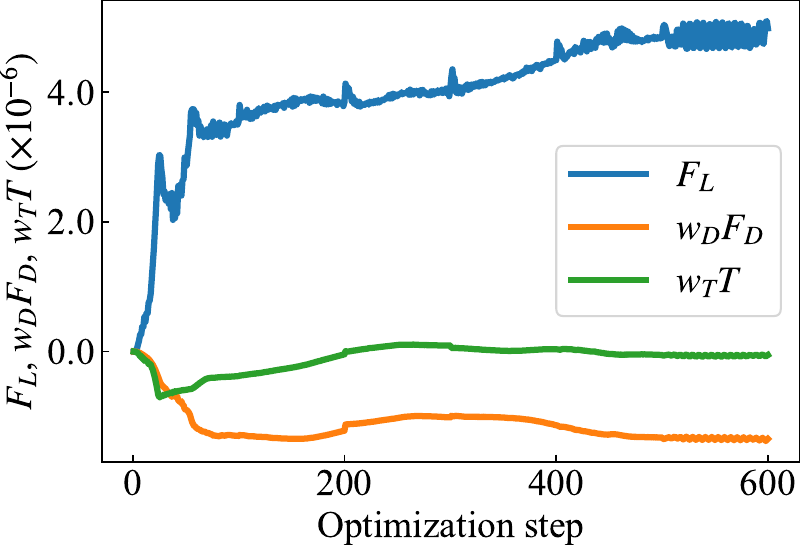}
        \subcaption{Breakdown of the objective functional}
        \label{fig:example2_history_breakdown}
    \end{minipage}
    \caption{Histories of the objective and constraint functional values and breakdown of the objective functional value for the \textit{2D oar}.}
\end{figure}
\begin{figure}
    \centering
    \begin{minipage}[t]{0.49\columnwidth}
        \centering
        \includegraphics[width=0.6\columnwidth]{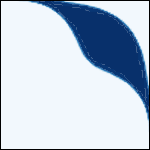}
        \subcaption{Optimized shape without the island}
        \label{fig:example2_mod_shape}
    \end{minipage}
    \begin{minipage}[t]{0.49\columnwidth}
        \centering
        \includegraphics[width=0.9\columnwidth]{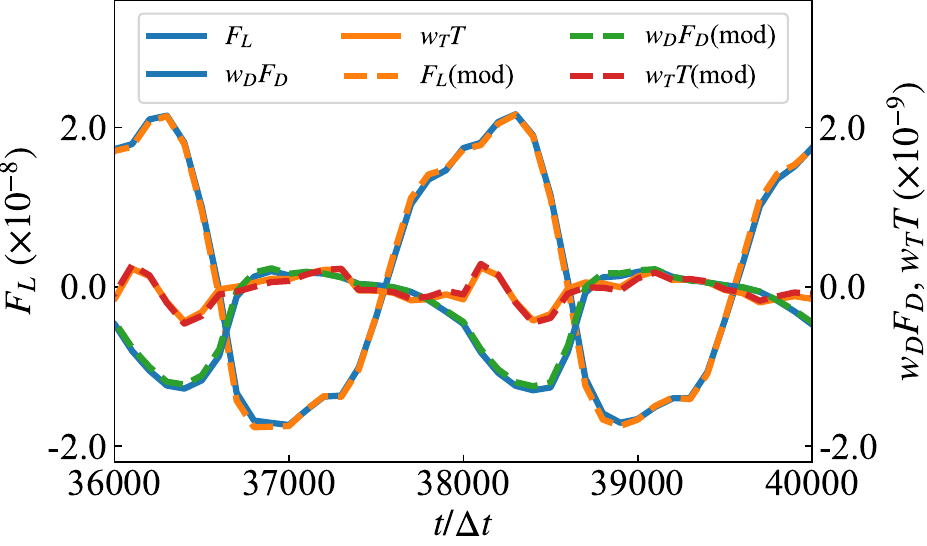}
        \subcaption{Time series of $F_L$, $w_DF_D$, and $w_TT$}
        \label{fig:example2_history_compare}
    \end{minipage}
    \caption{Optimized shape without the island and time series of $F_L$, $w_DF_D$, and $w_TT$ for the \textit{2D oar}.}
\end{figure}
\begin{figure}
    \centering
    \begin{minipage}[t]{0.28\columnwidth}
        \centering
        \includegraphics[width=\columnwidth]{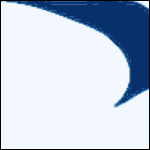}
        \subcaption{Optimized shape without diagonal symmetry}
        \label{fig:example2_optimized_shape_nosym}
    \end{minipage}
    \begin{minipage}[t]{0.7\columnwidth}
        \centering
        \includegraphics[width=0.95\columnwidth]{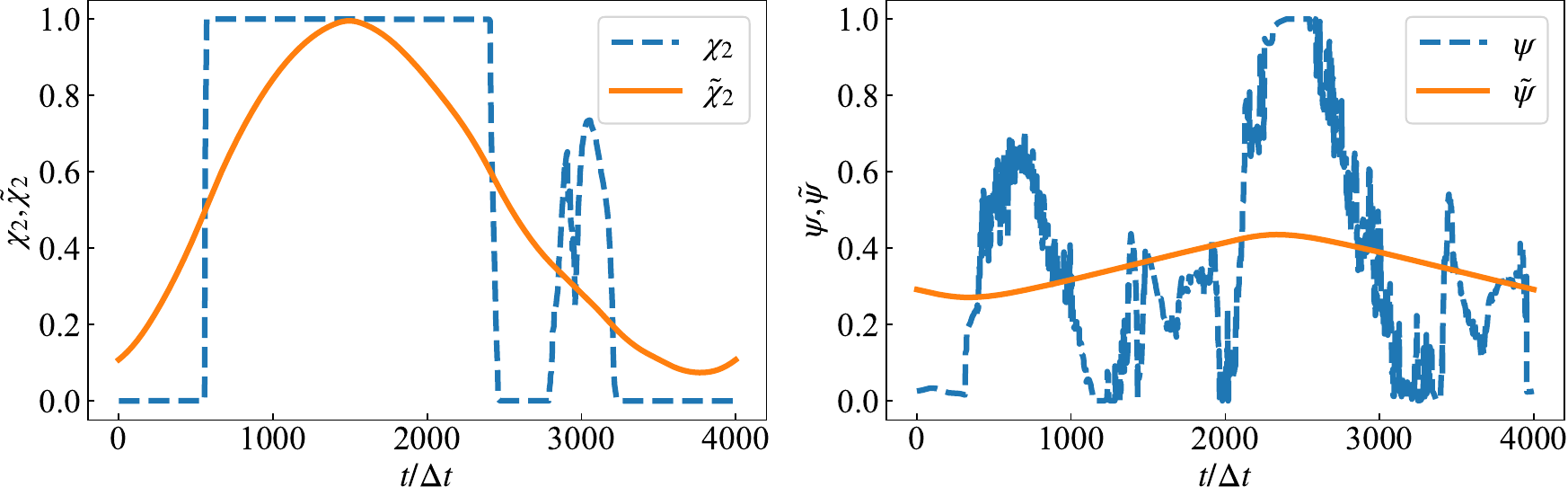}
        \subcaption{Optimized motion without diagonal symmetry}
        \label{fig:example2_optimized_motion_nosym}
    \end{minipage}
    \caption{Optimized shape and motion without diagonal symmetry for the \textit{2D oar}.}
\end{figure}
\begin{figure}
    \centering
    \includegraphics[width=0.5\columnwidth]{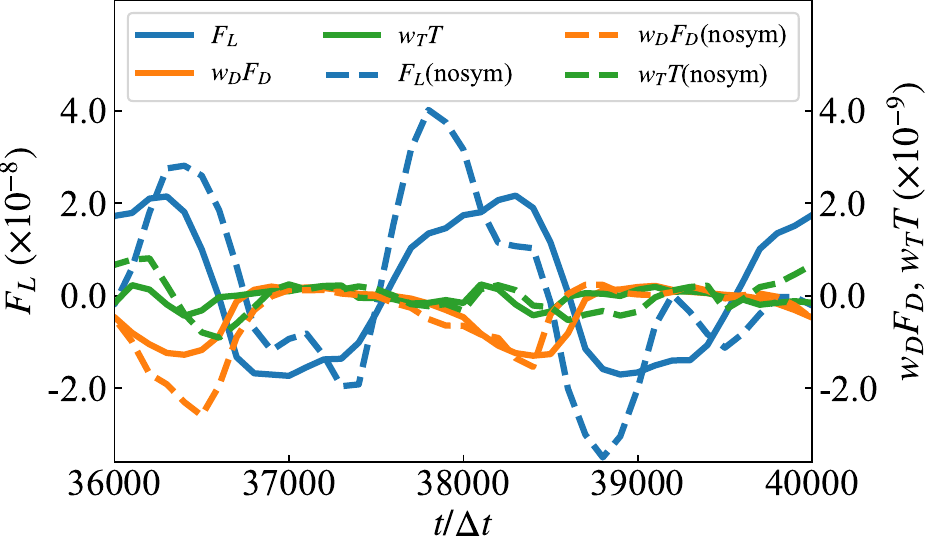}
    \caption{Time series of $F_L$, $w_DF_D$, and $w_TT$ for the optimized shapes with and without diagonal symmetry for the \textit{2D oar}.}
    \label{fig:example2_compare_nosym}
\end{figure}
\begin{figure}
    \centering
    \includegraphics[width=0.5\columnwidth]{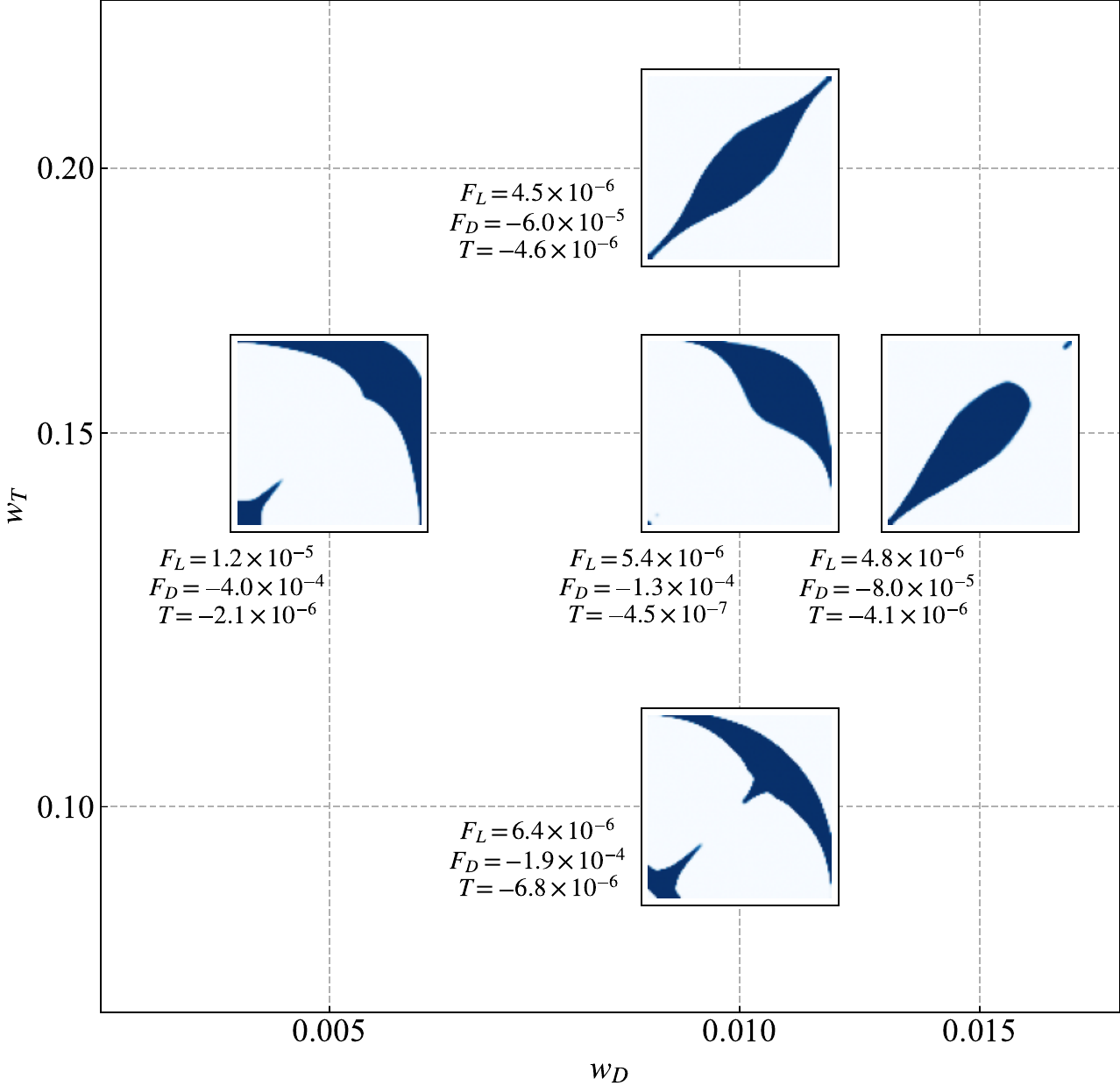}
    \caption{Optimized shapes for different combinations of $w_D$ and $w_T$ for the \textit{2D oar}.}
    \label{fig:example2_optimized_shape_compare}
\end{figure}

As a second example, we consider a two-dimensional oar design problem, referred to as a \textit{2D oar}.
This problem setting is inspired by what is called ``ro'' in Japanese, a type of paddle that was once widely used across Asia.
A characteristic feature of the ``ro'' is that it is operated with a single oar that remains in the water throughout the motion, in contrast to conventional paddles, which are typically used in pairs and alternate between strokes in water and recovery in air.
Based on this feature, the problem is formulated in a single-phase fluid, where the objective is to maximize the force exerted on the rigid body by optimizing both its shape and motion.

As shown in Fig.~\ref{fig:example2_design_setting}, the analysis domain $\mathcal{O}$ is a rectangle, and the design domain $D$ undergoes vertical translation and rotation; horizontal translation is not allowed.
The objective functional, $J_2$, is defined in Eq.~\eqref{eq:objective2} to maximize the horizontal force exerted by the fluid while minimizing vertical and rotational drag.
Both constraint functionals, $G_2$ defined in Eq.~\eqref{eq:constraint2} and $G_3$ defined in Eq.~\eqref{eq:constraint3}, are imposed.
Here, $G_2$ limits the maximum fluid volume; in other words, this constraint limits the minimum solid volume.
$G_2$ is applied to prevent the generation of excessively thin components.
$G_3$ is applied to suppress rigid-body velocities that are too large for the stability of the LBM-family method.
The other parameters are as follows: $N_\text{t}=4000\Delta t$, $\nu=0.1\Delta x$, $N_\text{opt}=100$, $q=0.1$, $\kappa^\text{ref}_\text{max}=300$, $\beta=\left\{1,2,4,8,16,32\right\}$, $R_\text{t}=1000\Delta t$, $R_\omega=2000\Delta t$, $R=4.8\Delta x$, $p_\omega=16$, $\omega_\text{max}=0.001$, $V_\text{max}=0.8$, and $w_D=0.01$ and $w_T=0.15$.
$l_m$ is set to half the diagonal length of the design grid.
As in the previous example, the periodic approximation is employed.
Additionally, we impose the diagonal symmetry condition for the shape.
The initial shape is given uniformly by $\gamma=0.5$, and the initial motion is given uniformly by $\chi_2=0.5$ and $\psi=0.5$.
The initial condition of the fluid is $p=1/3$ and $u_\alpha=0$.

The optimized shape is shown in Fig.~\ref{fig:example2_optimized_shape}, and the optimized motion is shown in Fig.~\ref{fig:example2_optimized_motion}.
The optimized shape looks like a crescent moon, and both the vertical translational motion and rotational motion are almost symmetric.
Additionally, the pressure distributions with flow velocity vector plots at each time step are shown in Figs.~\ref{fig:example2_pressure_at_each_time}.
The optimized shape repeats up-and-down motion while tilting to push the fluid, and is receives the reaction force from the fluid flow.
The overall shape is smooth and is suitable for reducing drag force and drag torque.
The history of the objective and constraint functionals are plotted in Fig.~\ref{fig:example2_history}, and the breakdown of the objective functional, $F_L$, $F_D$, and $T$, is plotted in Fig.~\ref{fig:example2_history_breakdown}.
Although the objective functional value $J_2$ shows an overall increasing trend, it exhibits oscillatory behavior.
This is likely because, in the \textit{2D oar} problem, three objective terms are considered in a weighted sum, two constraint functionals are imposed, and a periodic approximation is employed; therefore, the optimization problem becomes more complex than the \textit{2D pump} problem.
There is a small floating island, and we examine its effect by comparing the objective functional value with that for the optimized shape without the island.
The optimized shape without the island, which was modified by hand, is shown in Fig.~\ref{fig:example2_mod_shape}.
The comparison result is shown in Fig.~\ref{fig:example2_history_compare}, and the general trends are similar to each other; therefore, in practice, the island should be removed from the perspective of manufacturing.

We examine the effect of imposing the diagonal symmetry condition on the shape.
We compare the optimized shape obtained with the diagonal symmetry condition with that obtained without any symmetry condition.
The optimized shapes and motions are shown in Figs.~\ref{fig:example2_optimized_shape_nosym} and \ref{fig:example2_optimized_motion_nosym}.
Additionally, the breakdowns of the objective functional values for each case, computed a posteriori, are plotted in Fig.~\ref{fig:example2_compare_nosym}.
As shown in Fig.~\ref{fig:example2_compare_nosym}, the case without the diagonal symmetry condition has a larger thrust at the peak; however, the maximum pushed-back force from the fluid flow, corresponding to the negative thrust, is also larger.
In contrast, the case with the diagonal symmetry condition has relatively smaller thrust; however, the pushed-back forces are also suppressed.
Therefore, although the suitable condition depends on the design philosophy, the diagonal symmetry condition is more effective when smaller thrust fluctuation is desired.

We also examine the effects of the weighting factors $w_D$ and $w_T$.
The optimized shapes obtained for different combinations of $w_D$ and $w_T$ are shown in Fig.~\ref{fig:example2_optimized_shape_compare}.
The corresponding values of $F_L$, $F_D$ and $T$, evaluated a posteriori for each optimized shape, are also shown in the figure.
As shown in Fig.~\ref{fig:example2_optimized_shape_compare}, when either $w_D$ or $w_T$ is small, the optimized shape consists of a crescent-shaped component in the upper-right region and an isolated island in the lower-left region.
The crescent-shaped component appears to deflect the flow and thereby generate reaction forces, while the isolated island narrows the flow passage and accelerates the flow.
However, this configuration also tends to increase drag.
In contrast, when either $w_D$ or $w_T$ is increased, the optimized shape becomes more needle-like.
This may be because the contributions of $F_D$ and $T$ to the objective functional become more significant, and the optimizer suppress drag force and drag torque by avoiding the crescent-shape with the isolated island.
This trend is clearly observed for variations in $w_D$: as $w_D$ increases, $F_D$ decreases.
However, the same trend is not clearly observed for variation in $w_T$.
In particular, $T$ becomes very small when $w_T=0.15$.
This may be because of the complexity of the present optimization problem, which involves a weighted-sum multi-objective functional, multiple constraints, and a periodic approximation.
As a result, the optimization landscape may contain many local optima, and the case with $w_D=0.01$ and $w_T=0.15$ may have converged to a particularly favorable local optimum.

\subsection{3D pump}
\label{sec43}

\begin{figure}[t]
    \centering
    \begin{minipage}[t]{0.33\columnwidth}
        \centering
        \includegraphics[width=0.9\columnwidth]{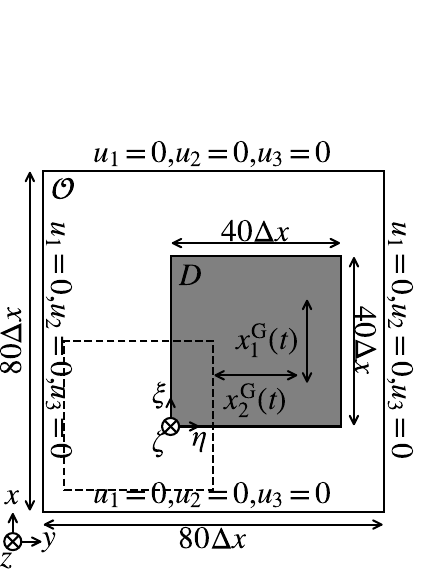}
        \subcaption{Bottom view}
        \label{fig:example3_design_setting_bottom}
    \end{minipage}
    \begin{minipage}[t]{0.33\columnwidth}
        \centering
        \includegraphics[width=0.9\columnwidth]{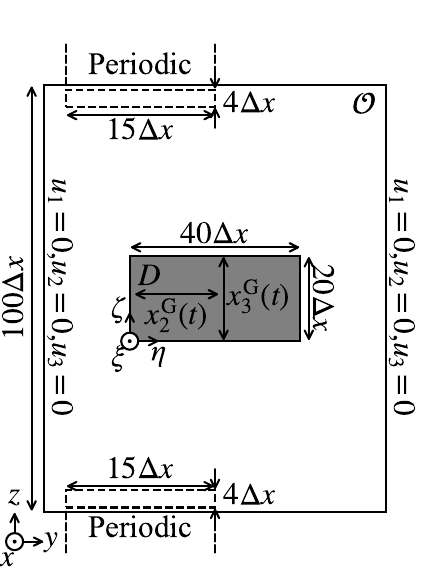}
        \subcaption{Front view}
        \label{fig:example3_design_setting_front}
    \end{minipage}
    \begin{minipage}[t]{0.33\columnwidth}
        \centering
        \includegraphics[width=0.9\columnwidth]{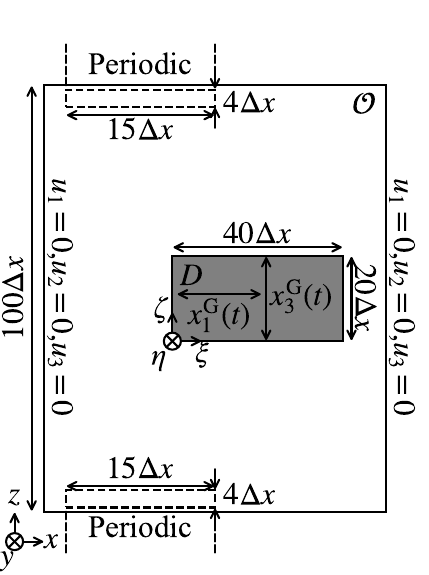}
        \subcaption{Side view}
        \label{fig:example3_design_setting_side}
    \end{minipage}
    \caption{Design setting for the \textit{3D pump}.}
\end{figure}
\begin{figure}[t]
    \centering
    \begin{minipage}[t]{0.33\columnwidth}
        \centering
        \includegraphics[width=0.9\columnwidth]{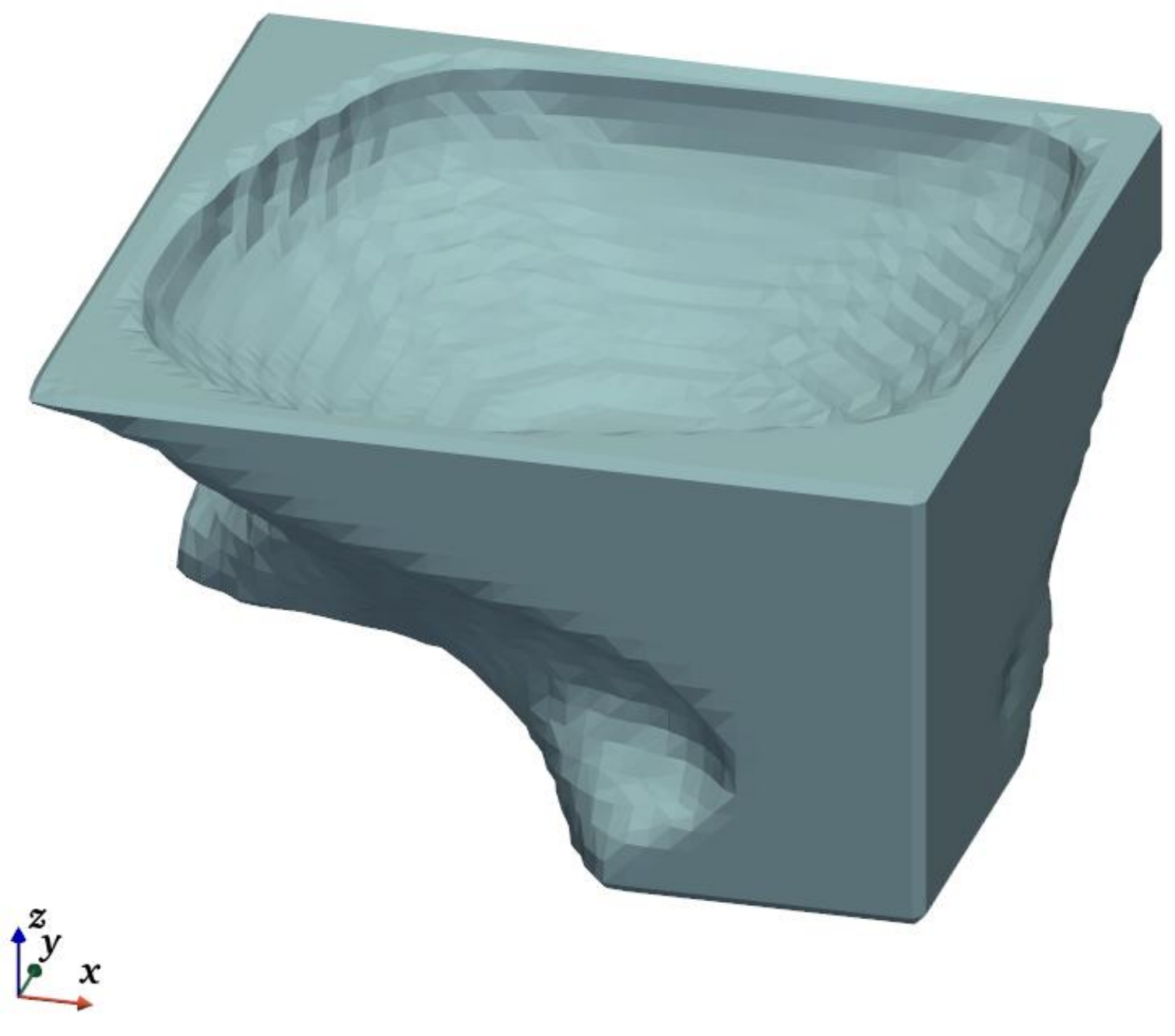}
        \subcaption{Isometric view}
        \label{fig:example3_optimized_shape_iso}
    \end{minipage}
    \begin{minipage}[t]{0.33\columnwidth}
        \centering
        \includegraphics[width=0.9\columnwidth]{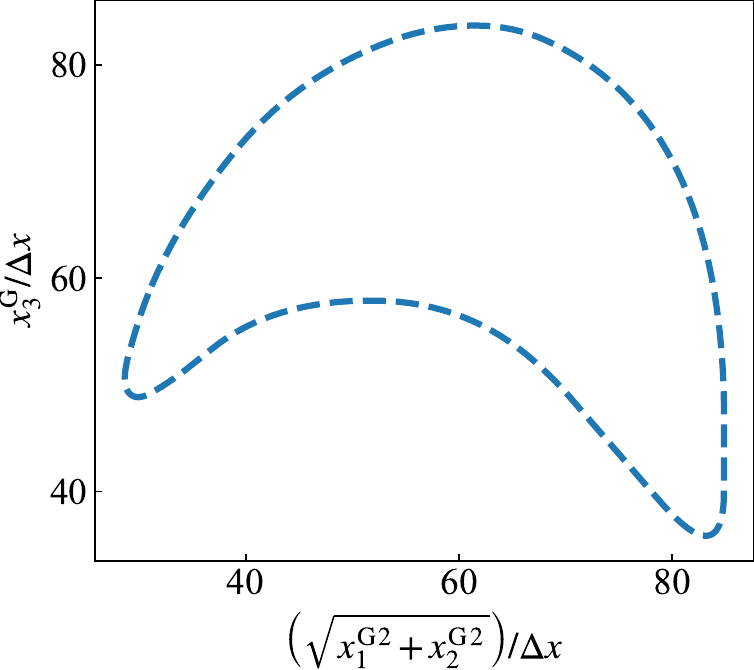}
        \subcaption{Optimized trajectory}
        \label{fig:example3_optimized_trajectory}
    \end{minipage} \\
    \begin{minipage}[t]{0.33\columnwidth}
        \centering
        \includegraphics[width=0.9\columnwidth]{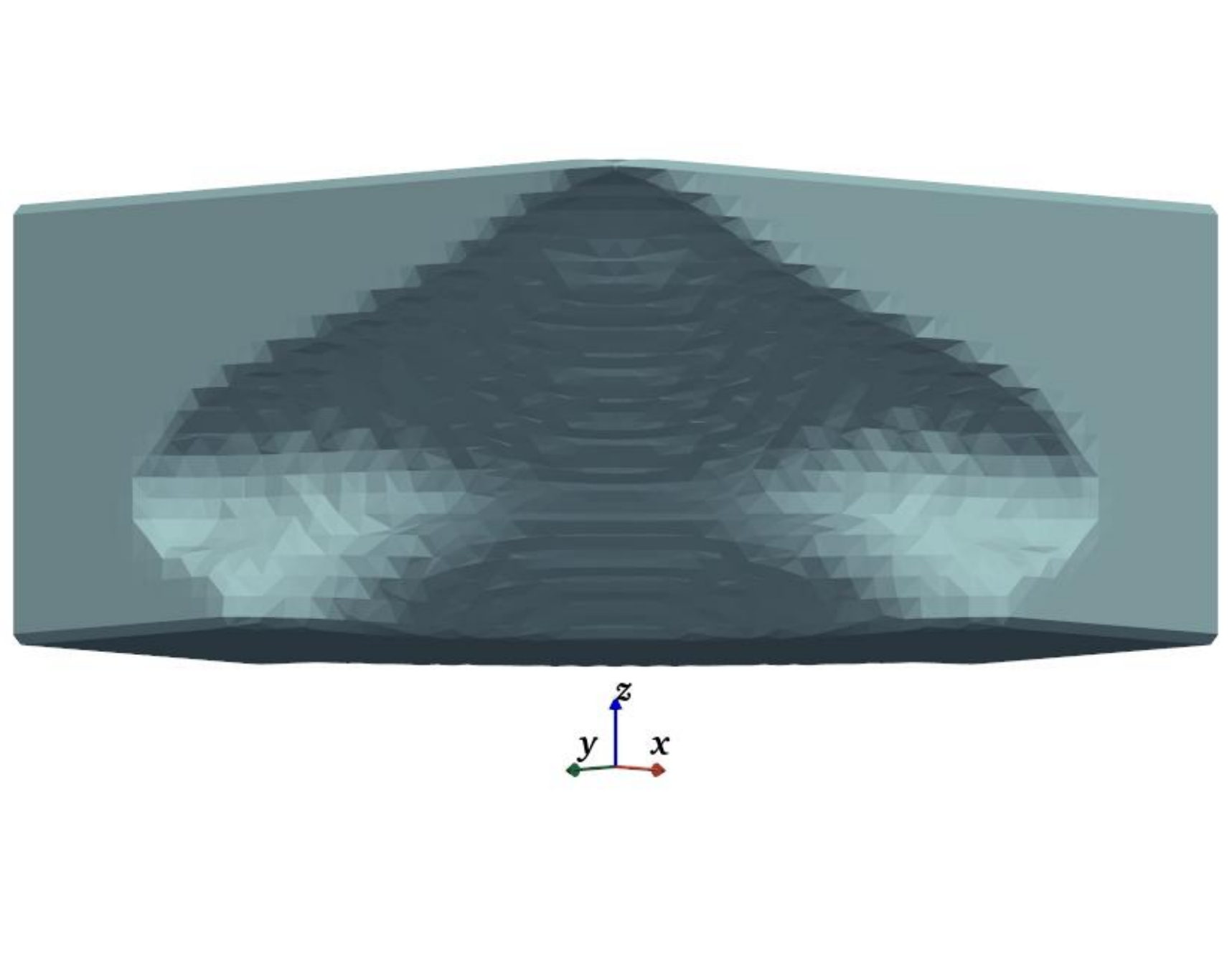}
        \subcaption{Front view}
        \label{fig:example3_optimized_shape_fron}
    \end{minipage}
    \begin{minipage}[t]{0.33\columnwidth}
        \centering
        \includegraphics[width=0.9\columnwidth]{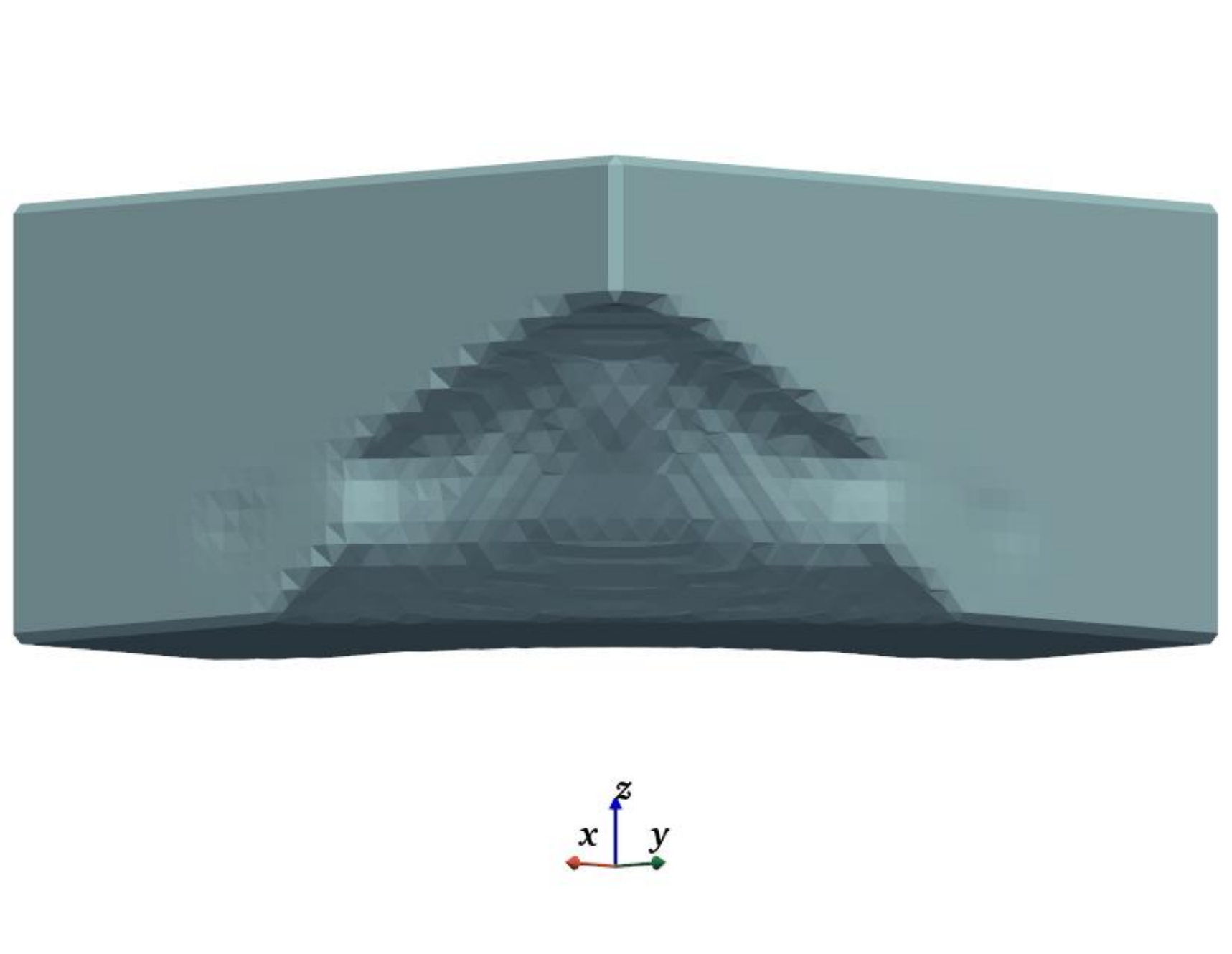}
        \subcaption{Back view}
        \label{fig:example3_optimized_shape_back}
    \end{minipage}
    \begin{minipage}[t]{0.33\columnwidth}
        \centering
        \includegraphics[width=0.9\columnwidth]{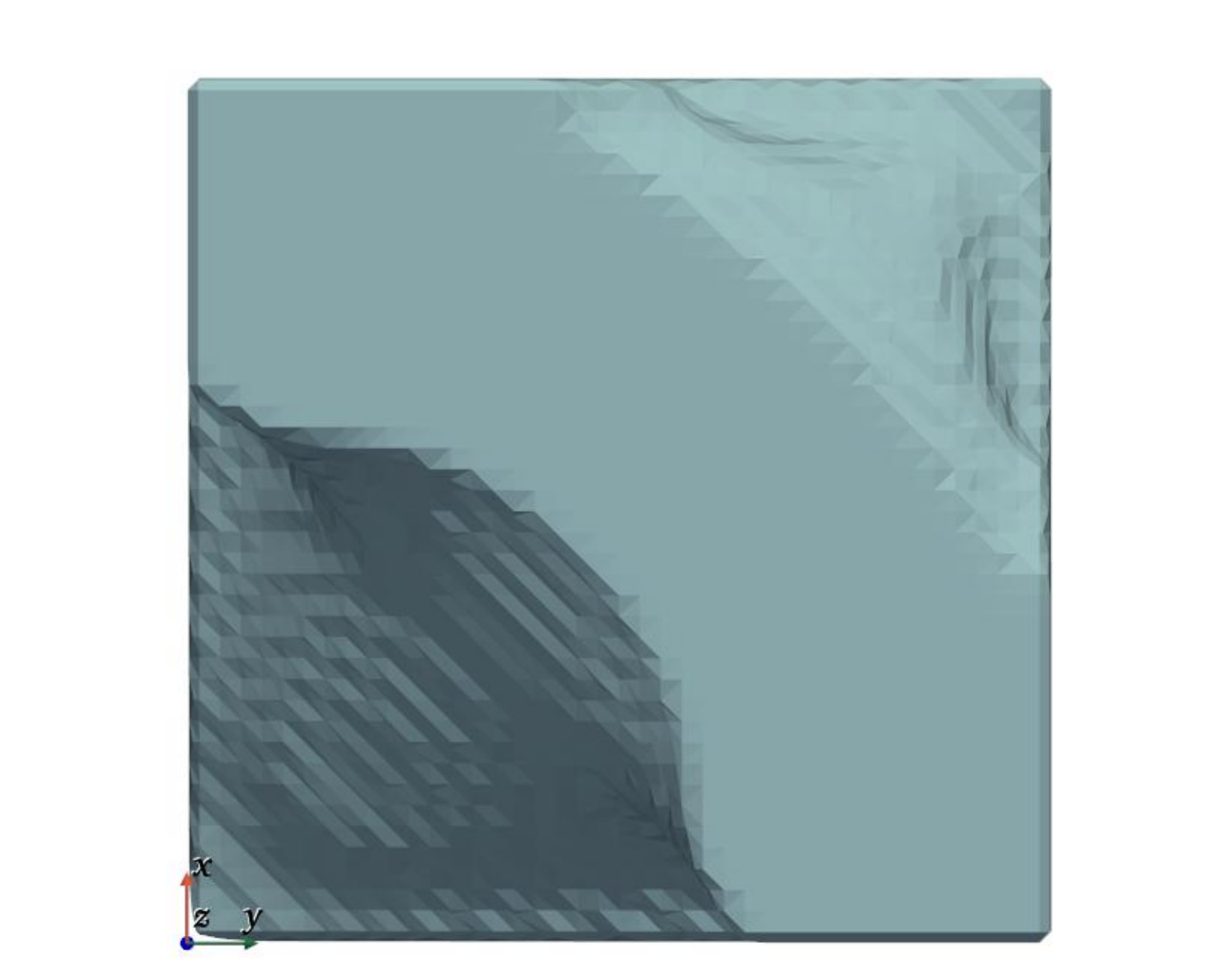}
        \subcaption{Bottom view}
        \label{fig:example3_optimized_shape_bottom}
    \end{minipage} \\
    \vspace{0.5em}
    \begin{minipage}[t]{\columnwidth}
        \centering
        \includegraphics[width=0.9\columnwidth]{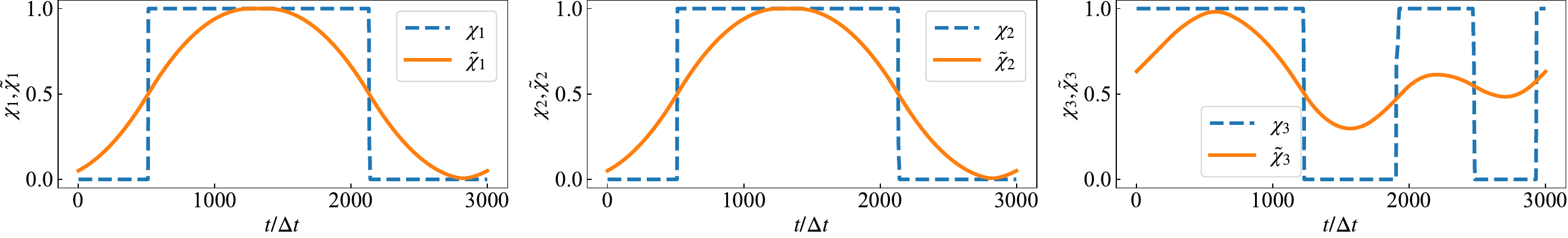}
        \subcaption{Optimized motion}
        \label{fig:example3_optimized_motion}
    \end{minipage}
    \caption{Optimized shape, trajectory and motion for the \textit{3D pump}.}
\end{figure}
\begin{figure}[t]
    \centering
    \begin{minipage}[t]{0.49\columnwidth}
        \centering
        \includegraphics[width=0.9\columnwidth]{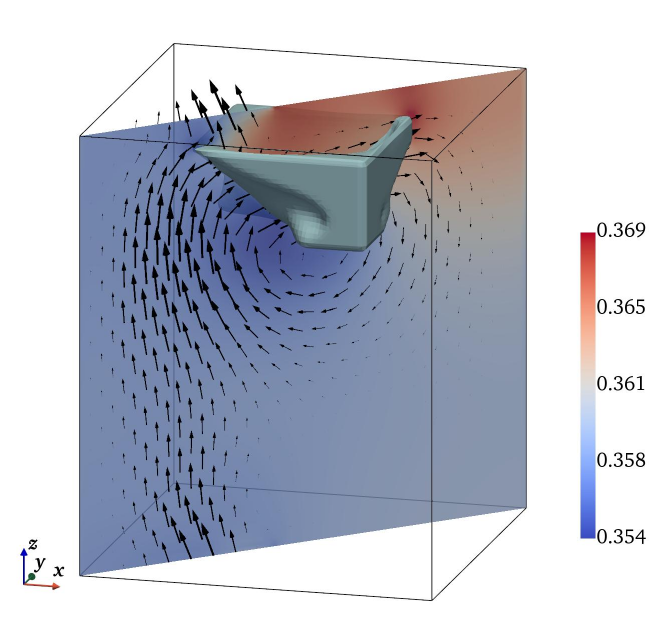}
        \subcaption{$t/\Delta t=500$}
        \label{fig:example3_pressure_t_500}
    \end{minipage}
    \begin{minipage}[t]{0.49\columnwidth}
        \centering
        \includegraphics[width=0.9\columnwidth]{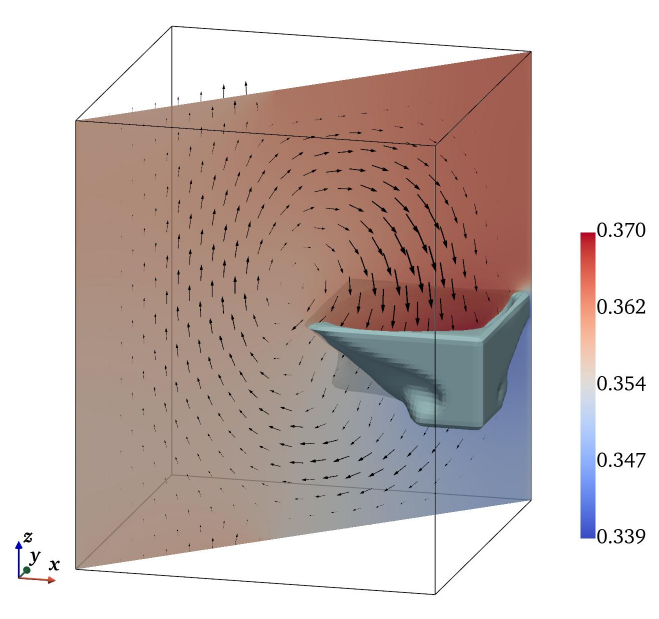}
        \subcaption{$t/\Delta t=1500$}
        \label{fig:example3_pressure_t_1500}
    \end{minipage} \\
    \begin{minipage}[t]{0.49\columnwidth}
        \centering
        \includegraphics[width=0.9\columnwidth]{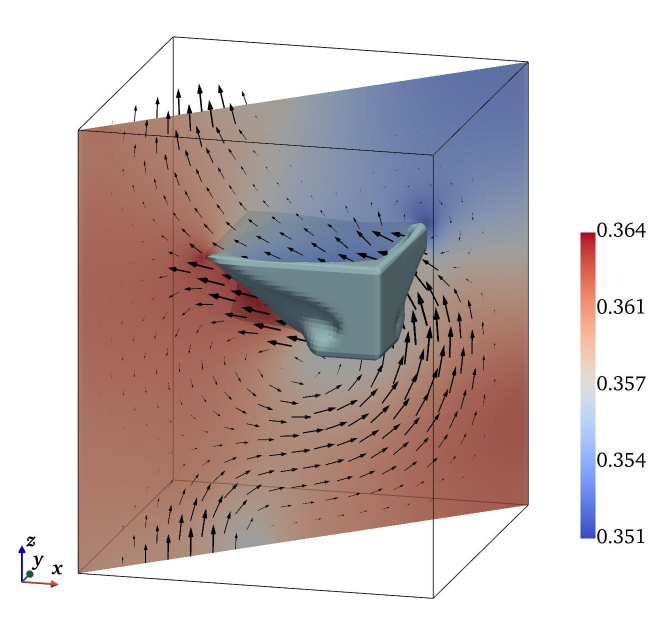}
        \subcaption{$t/\Delta t=2100$}
        \label{fig:example3_pressure_t_2100}
    \end{minipage}
    \begin{minipage}[t]{0.49\columnwidth}
        \centering
        \includegraphics[width=0.9\columnwidth]{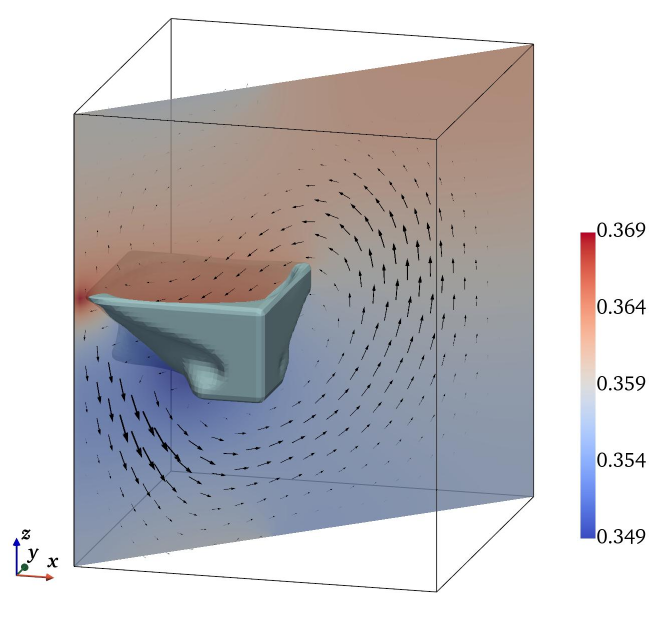}
        \subcaption{$t/\Delta t=2700$}
        \label{fig:example3_pressure_t_2700}
    \end{minipage}
    \caption{Pressure distributions and flow velocity vector plots at each time step for optimized shape and motion of the \textit{3D pump}.}
\end{figure}
\begin{figure}[t]
    \centering
    \begin{minipage}[t]{0.33\columnwidth}
        \centering
        \includegraphics[width=0.9\columnwidth]{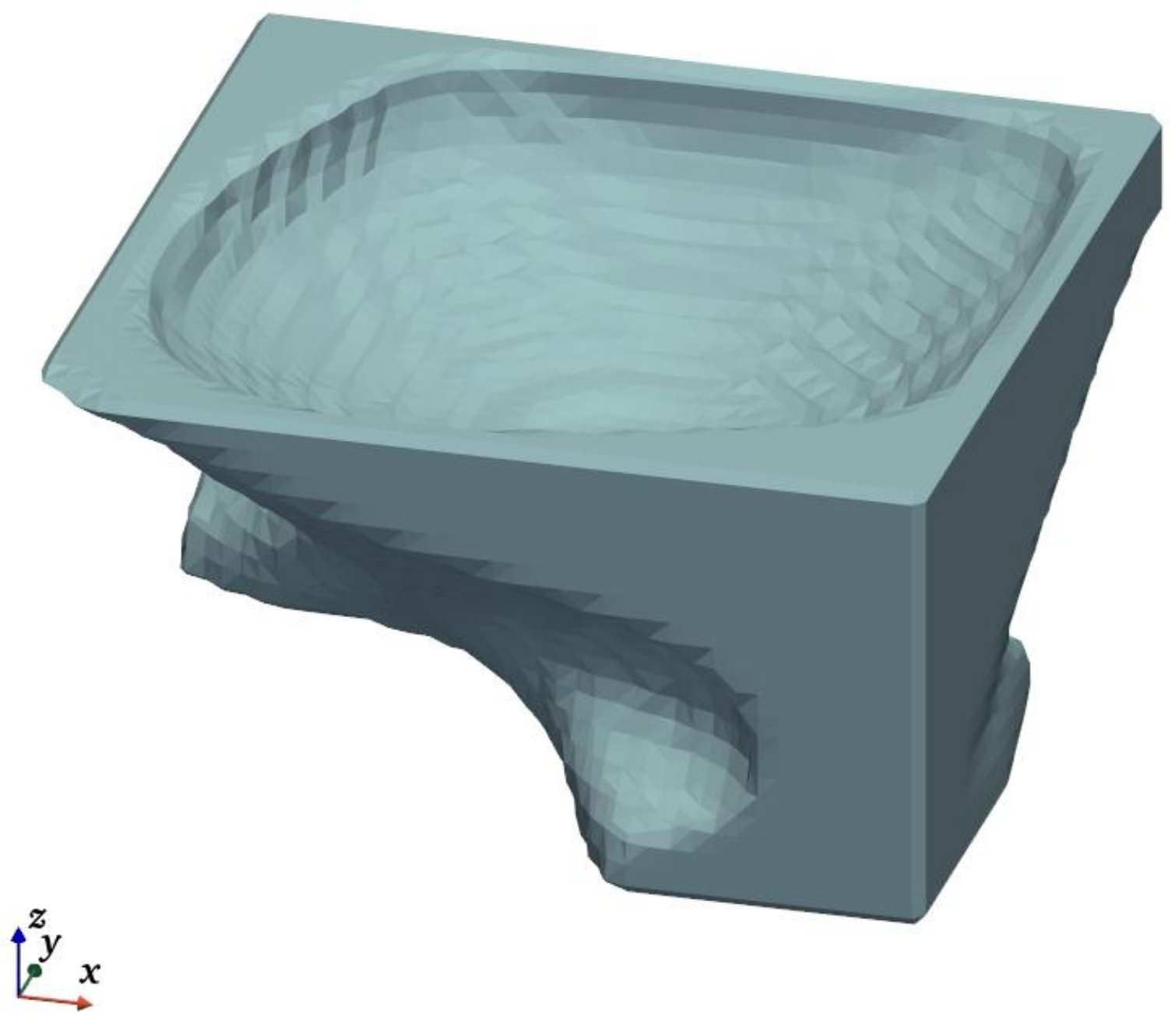}
        \subcaption{Optimized shape}
        \label{fig:example3_optimized_shape_iso_v03}
    \end{minipage}
    \begin{minipage}[t]{0.33\columnwidth}
        \centering
        \includegraphics[width=0.9\columnwidth]{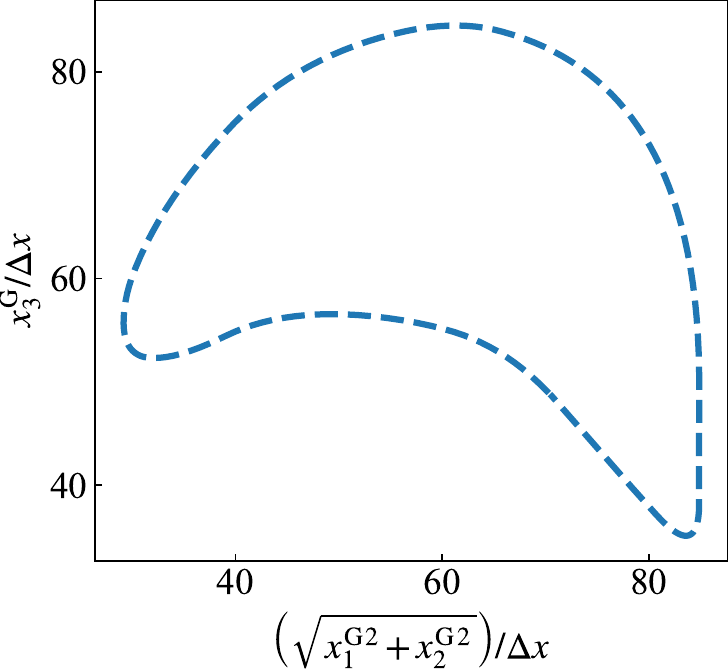}
        \subcaption{Optimized trajectory}
        \label{fig:example3_optimized_trajectory_v03}
    \end{minipage} \\
    \begin{minipage}[t]{0.33\columnwidth}
        \centering
        \includegraphics[width=0.9\columnwidth]{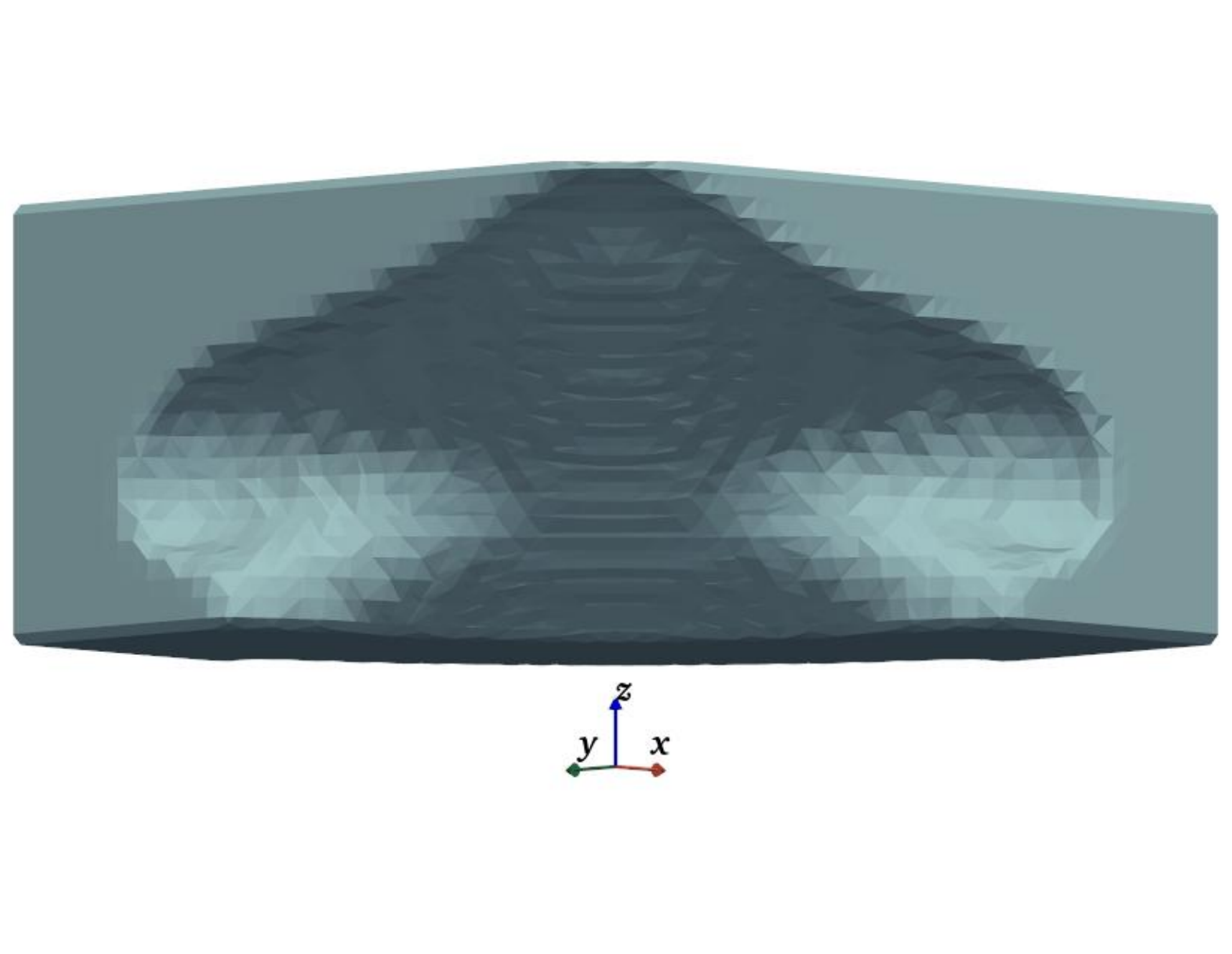}
        \subcaption{Front view}
        \label{fig:example3_optimized_shape_fron_v03}
    \end{minipage}
    \begin{minipage}[t]{0.33\columnwidth}
        \centering
        \includegraphics[width=0.9\columnwidth]{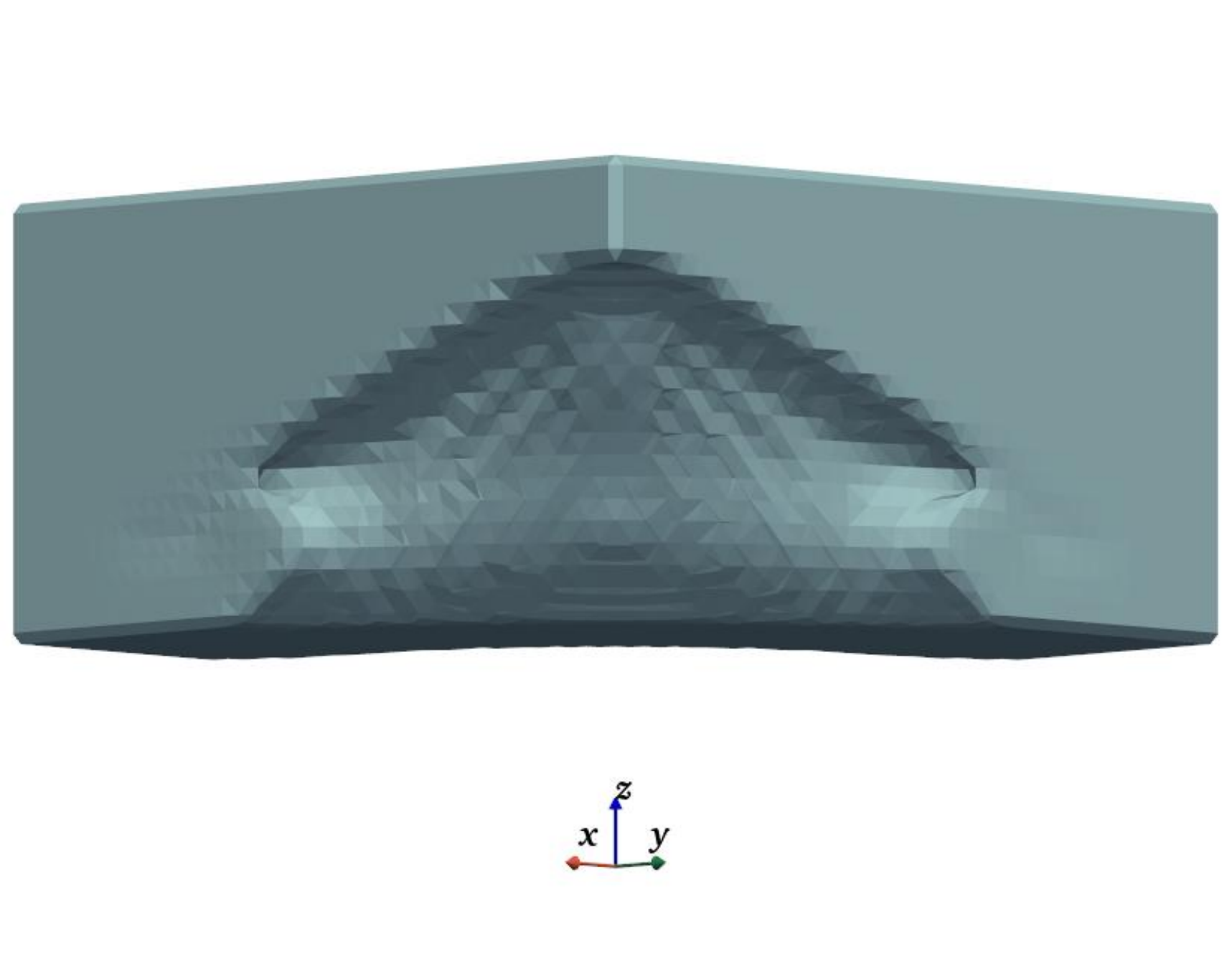}
        \subcaption{Back view}
        \label{fig:example3_optimized_shape_back_v03}
    \end{minipage}
    \begin{minipage}[t]{0.33\columnwidth}
        \centering
        \includegraphics[width=0.9\columnwidth]{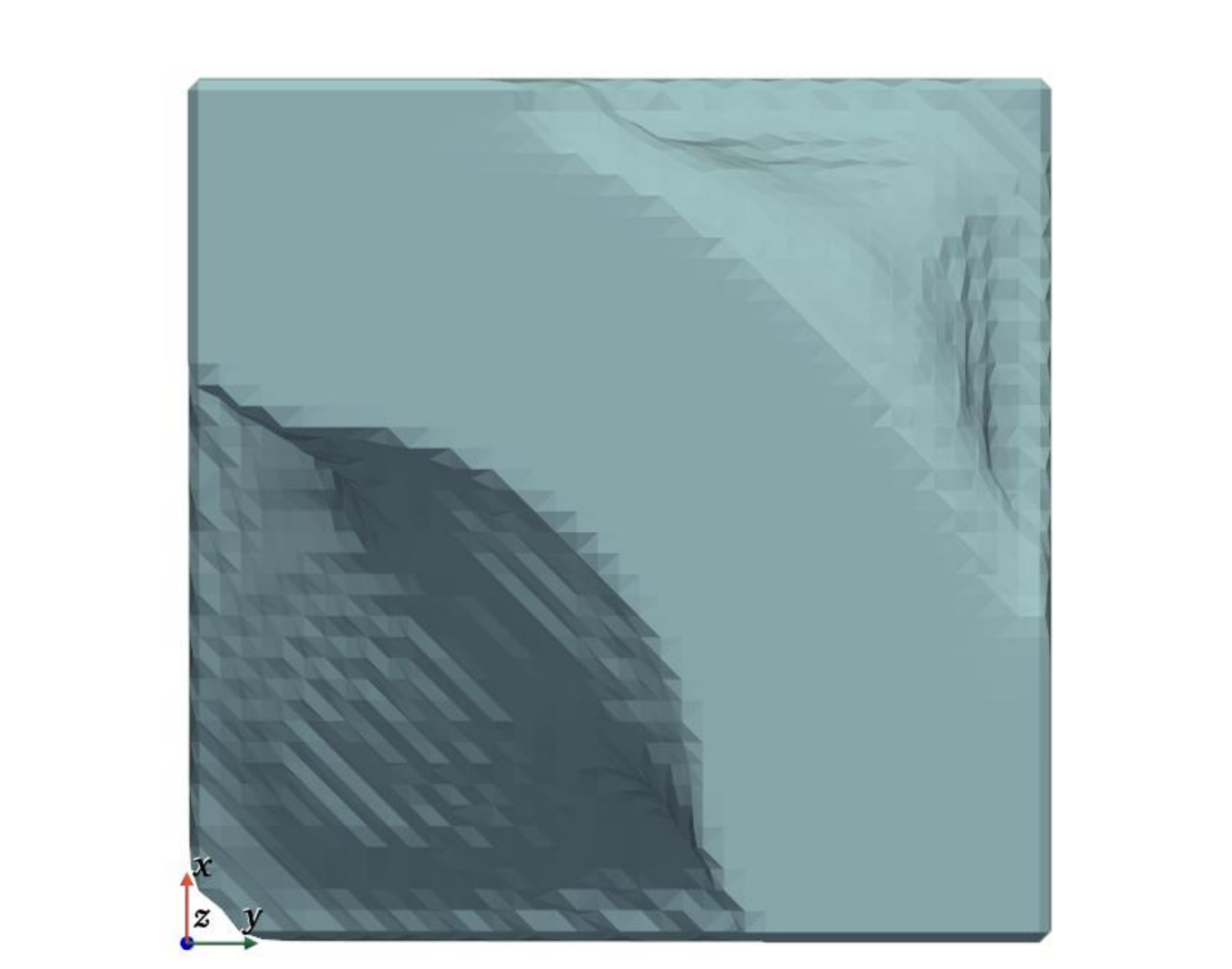}
        \subcaption{Bottom view}
        \label{fig:example3_optimized_shape_bottom_v03}
    \end{minipage} \\
    \vspace{0.5em}
    \begin{minipage}[t]{\columnwidth}
        \centering
        \includegraphics[width=0.9\columnwidth]{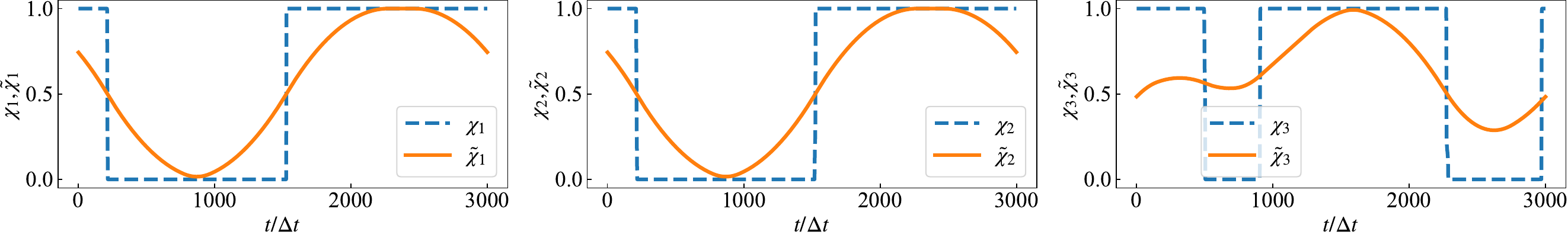}
        \subcaption{Optimized motion}
        \label{fig:example3_optimized_motion_v03}
    \end{minipage}
    \caption{Optimized shape, trajectory, and motion for the \textit{3D pump} with $\phi=0.3$.}
\end{figure}
\begin{figure}[t]
    \centering
    \includegraphics[width=0.5\columnwidth]{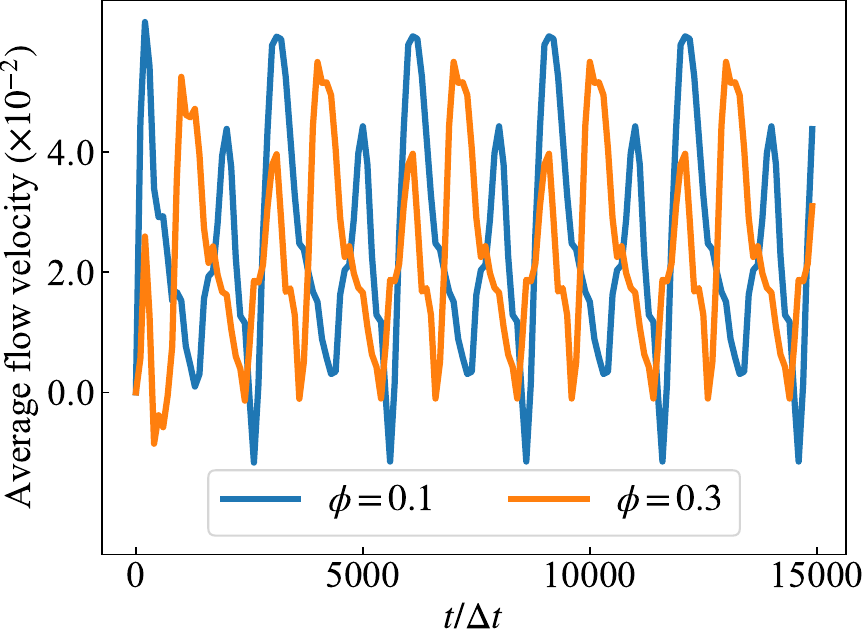}
    \caption{Comparison of the average flow velocity for the \textit{3D pump} with $\phi=0.1$ and $\phi=0.3$.}
    \label{fig:example3_comparison}
\end{figure}

As a final example, we consider the three-dimensional case of simultaneously designing the shape and translational motion, referred to as a \textit{3D pump}.
The design setting is shown in Figs.~\ref{fig:example3_design_setting_bottom}--\subref{fig:example3_design_setting_side}.
Wall boundaries are applied to all boundaries except for the square regions on the top and bottom walls, indicated by the dashed lines in Figs.~\ref{fig:example3_design_setting_bottom}--\subref{fig:example3_design_setting_side}.
The boundaries surrounded by the dashed lines are periodic boundaries.
The design region $D$ is allowed to undergo only translational motion and is prohibited from rotational motion; in other words, only Eq.~\eqref{eq:x_g} is considered, where $\alpha$ takes the values $1$, $2$, and $3$.
The objective functional is defined as the weighted sum of the average and variance of the flow velocity, as expressed in Eq.~\eqref{eq:objective1}.
The target flow direction is $\bm{n}=\left(0,0,1\right)$, and the target region $\mathcal{O}\text{target}$, where the flow velocity is optimized, is represented by the box enclosed with dashed lines in Figs.~\ref{fig:example3_design_setting_bottom}--\subref{fig:example3_design_setting_side}, located just below the periodic boundary.
The other parameters are as follows: $N\text{t}=3000\Delta t$, $\nu=0.1\Delta x$, $N_\text{opt}=100$, $\kappa^\text{ref}_\text{max}=100$, $q=0.1$, $\beta=\left\{1,2,4,8,16\right\}$, $R=4.8\Delta x$, and $R\text{t}=750\Delta t$.
Additionally, as in the previous examples, the periodic approximation described in Section~\ref{sec33} is employed to efficiently solve the periodic problem.
The initial shape is uniformly $\gamma=0.5$, and the initial motion is uniformly $\chi_\alpha=0.5$.
The initial condition of the fluid is $p=1/3$ and $u_\alpha=0$.

First, we perform the optimization with the weight for the flow velocity variance in Eq.~\eqref{eq:objective1} set to $\phi=0.1$, and the optimized shape, trajectory, and motion are shown in Figs.~\ref{fig:example3_optimized_shape_iso}--\subref{fig:example3_optimized_motion}, respectively.
As shown in Fig.~\ref{fig:example3_optimized_motion}, the motions in the $x$ and $y$ directions are almost identical; in other words, the rigid body moves only on the plane described by $x=y$.
It should be noted that, although no symmetric conditions, such as shape symmetry conditions, are imposed, both the design setting and the objective functional are symmetric with respect to $x=y$.
Therefore, the resulting motion may also become symmetric with respect to $x=y$.
According to the feature of the motion described above, the horizontal axis in Fig.~\ref{fig:example3_optimized_trajectory} represents the position on the $x=y$ plane, calculated as $\sqrt{x^2+y^2}$.
Additionally, the pressure distributions and flow velocity vector plots on the $x=y$ plane at each time step are shown in Figs.~\ref{fig:example3_pressure_t_500}--\subref{fig:example3_pressure_t_2700}.
The optimized shape resembles a thin bucket, and to transport the fluid, the rigid body repeats the following motion: (a) it moves toward the upper center to carry the fluid to the target region without blocking the periodic boundary by the rigid body itself ($2700\leq t/\Delta t\leq3000,0\leq t/\Delta t\leq 600$), (b) it moves downward to carry the fluid again; moreover, it does not move directly downward in order to avoid generating reversal flow ($600\leq t/\Delta t\leq 1600$), and (c) it moves to the region below the target region while preventing the reversal flow from becoming large ($1600\leq t/\Delta t\leq 2700$).
Here, the average computation time for one optimization iteration is about $312$ seconds when using a single NVIDIA GeForce RTX 4070 GPU.

Next, we examine the effect of the weight parameter for the variance in the objective functional and perform the optimization with the setting $\phi=0.3$.
With this setting, the effect of the variance in the objective functional is larger than that in the case treated in the previous section, where $\phi=0.1$.
The other parameters are the same as those in the case with $\phi=0.1$.
The optimized shape, trajectory, and motion are shown in Figs.~\ref{fig:example3_optimized_shape_iso_v03}--\subref{fig:example3_optimized_motion_v03}, respectively.
Although there are slight differences, the optimized shapes are almost the same in both cases.
In contrast, the optimized trajectory for the case with $\phi=0.3$ is flatter than that for the case with $\phi=0.1$ while the rigid body moves toward the region below the target region.
This may be because, if the trajectory is flatter in that interval, the reversal flow is reduced since the rigid body moves less downward; however, the upward moving distance also becomes shorter, and therefore the maximum upward flow becomes smaller.
The comparison of the volumetric average flow in the target region between the cases with $\phi=0.1$ and $\phi=0.3$ is shown in Fig.~\ref{fig:example3_comparison}.
Here, the flow velocities in Fig.~\ref{fig:example3_comparison} are calculated from the initial conditions for the fluid field, $p=1/3$ and $u_\alpha=0$, using the optimized shape and motion a posteriori with five repeated cycles.
In contrast, during the optimization, the state fields are calculated using the periodic approximation described in Section~\ref{sec33}; in other words, the initial condition for each cycle is given by the value at the last time step of the previous cycle with the shape obtained in the previous optimization, which is generally different from the current shape.
As shown in Fig.~\ref{fig:example3_comparison}, except for the transient phase ($0\leq t/\Delta t\leq 2000$), the variation in the case with $\phi=0.3$ is smaller, and the reversal flow is more suppressed than that in the case with $\phi=0.1$, although the maximum flow velocity becomes smaller.
This result indicates that the proposed optimization method is effective for different variance weights.
From the optimization results described above, namely that the optimized shapes are almost the same whereas the trajectories are different, it is suggested that the motion has a greater influence on the variance effect than the shape.
This may be because a bucket-like shape is effective to some extent in both cases. 

\section{Conclusion}
\label{sec5}

In this study, we proposed a method for the simultaneous optimization of the shape and motion of a moving rigid body in fluid.
The state field was solved with the LKS, which is an extended version of the LBM.
The moving rigid body was represented with the grid separation approach, in which the design grid was separated from the analysis grid.
Not only the shape but also the rigid-body motion was optimized; therefore, the design variables consisted of the shape (i.e., the pseudo density) and the motion, which corresponded to the rigid body position at each time step.
The sensitivity analysis method was developed based on the adjoint variable method, and was used with optimization techniques such as the filtering scheme, the continuation scheme, and the periodic problem approximation for numerical examples in both two- and three-dimensional cases.

We studied two two-dimensional cases; one was referred to as the \textit{2D pump} and the other was referred to as the \textit{2D oar}, and in both cases, optimized shapes and motions that were strongly coupled with each other were obtained.
In the \textit{2D pump}, we examined the effectiveness of considering the motion as an optimization variable, and the optimized shape and motion were better than the cases considering only shape optimization or motion optimization.
We also examined the effects of the time interval $N_\text{t}$ and the maximum volume ratio $V_\text{max}$.
For $N_\text{t}$, similar optimized shapes and trajectories were observed repeatedly as $N_\text{t}$ was increased, indicating that there is a suitable range of $N_\text{t}$.
For $V_\text{max}$, the depth of the cavity on the upper surface of the optimized shape, which is effective for transporting fluid, differed depending on $V_\text{max}$.
However, the optimized trajectories were quite similar, indicating that the volume constraint mainly affects the representation of the shape.
In the \textit{2D oar}, we employed an objective functional defined as thrust weighted together with the translational and rotational resistances.
We examined the effect of imposing diagonal symmetry on the shape and showed that the optimized shape with diagonal symmetry suppressed thrust fluctuations more effectively than that without diagonal symmetry.
We also examined the effects of the weighting factors for drag force and drag torque, $w_D$ and $w_T$, in the objective functional, and showed the general trend of the optimized shape for each combination of $w_D$ and $w_T$.
An inconsistent trend was observed for $w_T$, which may be attributed to the complexity of the problem setting of the \textit{2D oar}.
We studied one three-dimensional case referred to as \textit{3D pump} and showed that the proposed method was applicable not only to two-dimensional cases but also to a three-dimensional one.
In \textit{3D pump}, we employed a weighted-sum objective functional, and showed that the proposed method optimized the shape and motion according to the weights. While the optimized shapes showed only minor differences for each weight value, the optimized motions exhibited clear differences depending on the weight values.

We represented the motion as the position of the rigid body at each time step; however, there were at least two other ways: one was velocity and the other was acceleration (or the control force).
We chose the position mainly due to its simplicity; in other words, position optimization did not require a constraint functional to forbid the rigid body from going out of the analysis region.
However, in realistic cases, optimizing the control force is more natural; therefore, in the future, we will tackle force optimization. 
\section{Replication of results}
The necessary information for replication of the results is presented in this paper.
The interested reader may contact the corresponding author for further implementation details.

\section*{Acknowledgements}
This work was supported by JSPS KAKENHI (GrantNo. 23K26018).

\appendix

\section{Verification of sensitivity analysis}
\label{appendixA}

\begin{figure}[t]
    \centering
    \begin{minipage}[t]{0.2\columnwidth}
        \centering
        \includegraphics[width=\columnwidth]{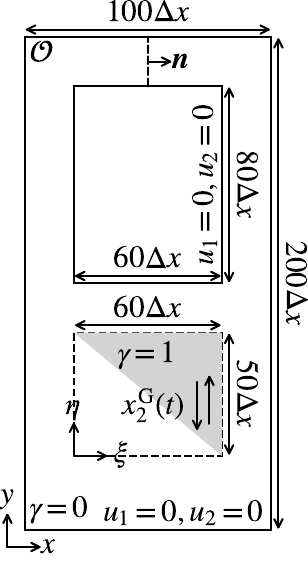}
        \subcaption{Problem setting}
        \label{fig:appendixA_problem_setting}
    \end{minipage}
    \begin{minipage}[t]{0.39\columnwidth}
        \centering
        \includegraphics[width=0.95\columnwidth]{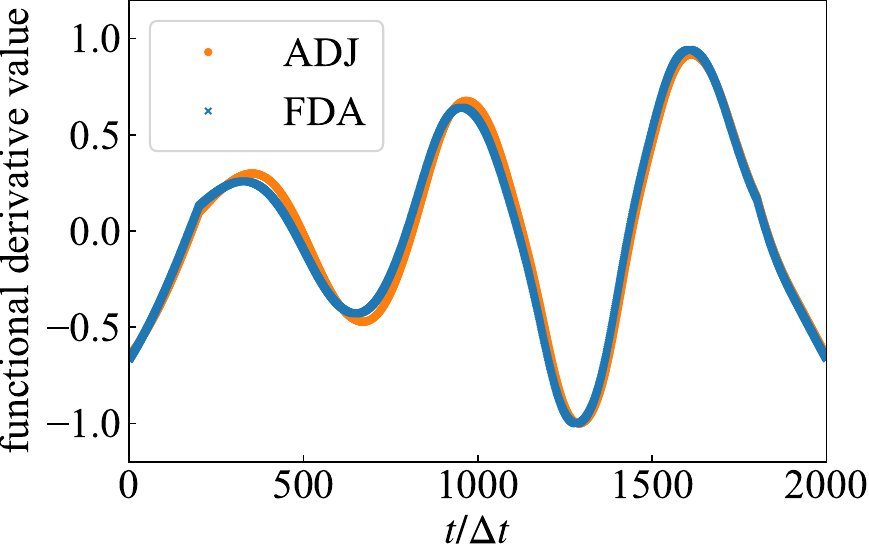}
        \subcaption{$\chi_2=0.5$}
        \label{fig:appendixA_sensitivity_const}
    \end{minipage}
    \begin{minipage}[t]{0.39\columnwidth}
        \centering
        \includegraphics[width=0.95\columnwidth]{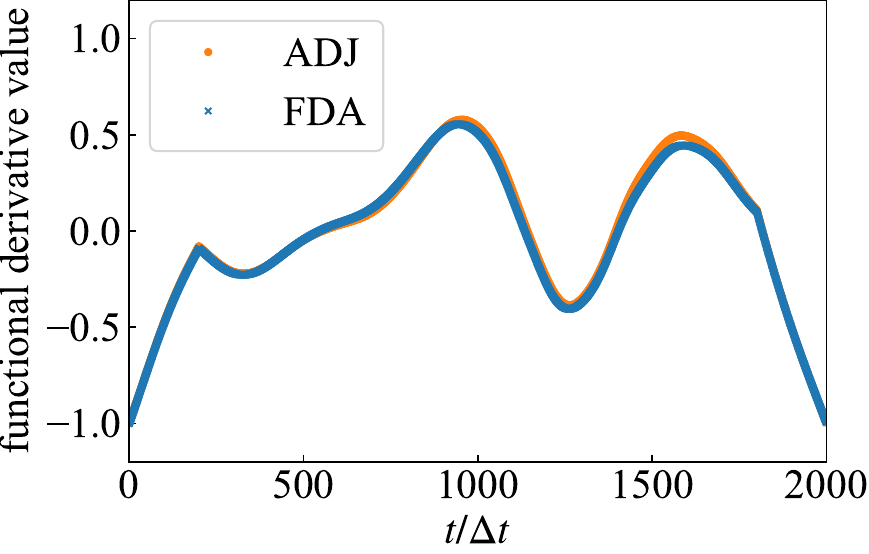}
        \subcaption{$\chi_2=0.9\times0.5\left\{\cos{\left(2\pi t/N_\text{t}\right)}+1\right\}$}
        \label{fig:appendixA_sensitivity_cos}
    \end{minipage}
    \caption{Problem setting and results of the design sensitivity verification.}
\end{figure}

In this section, we verify the design sensitivity for the motion calculated by the adjoint variable method by comparing it with that obtained by the Finite Difference Approximation (FDA).
The problem setting is shown in Fig.~\ref{fig:appendixA_problem_setting}, and all boundaries are imposed as wall boundaries.
The inner wall is located in the analysis domain, and the functional used to compute the design sensitivity is the average flow velocity defined in Eq.~\eqref{eq:objective1} with $\phi=0$, where the target region $\mathcal{O}\text{target}$ is located on the dashed line in Fig.~\ref{fig:appendixA_problem_setting}.
The rigid body is allowed to move only vertically and is prohibited from moving horizontally or rotationally.
The parameters are as follows: $N_\text{t}=2000\Delta t$, $\nu=0.1\Delta x$, $q=0.1$, $\kappa^\text{ref}_\text{max}=200$, $R_t=200\Delta t$, $x^\text{G}_{\text{2,max}}=75\Delta x$, and $x^\text{G}_{\text{2,min}}=25\Delta x$.
The initial conditions for the fluid are given by $p=1/3$ and $u_\alpha=0$.
We consider two types of motion: (a) $\chi_2=0.5$, i.e., the rigid body is stationary, and (b) $\chi_2=0.9\times0.5\left\{\cos{\left(2\pi t/N_\text{t}\right)}+1\right\}$.
The comparison results for each case are shown in Figs.~\ref{fig:appendixA_sensitivity_const} and \ref{fig:appendixA_sensitivity_cos}, respectively.
Here, ``ADJ'' in Figs.~\ref{fig:appendixA_sensitivity_const} and \ref{fig:appendixA_sensitivity_cos} denotes the design sensitivity obtained by the adjoint variable method.
As shown in Figs.~\ref{fig:appendixA_sensitivity_const} and \ref{fig:appendixA_sensitivity_cos}, although slight differences are observed, the results in both cases practically agree with each other. 
\section*{Conflict of interest}
The authors declare that they have no conflict of interest.

\end{document}